\documentclass[%
 reprint,
superscriptaddress,
 amsmath,amssymb,
 aps,
]{revtex4-2}

\usepackage{sidecap}
\usepackage{lipsum}  
\usepackage{graphicx}
\usepackage{dcolumn}
\usepackage{bm}
\usepackage{scalerel}
\usepackage{xcolor}

\begin{document}

\preprint{APS/123-QED}

\title{Method for Extracting the Equivalent Admittance from Time-Varying Metasurfaces and Its Application to Self-Tuned Spatiotemporal Wave Manipulation}

\author{Ashif Aminulloh Fathnan}
\email{fathnan.aminulloh@nitech.ac.jp}
\affiliation{Department of Engineering, Graduate School of Engineering, \\ Nagoya Institute of Technology, Nagoya, Aichi, 466-8555, Japan}
\author{Haruki Homma}
\affiliation{Department of Engineering, Graduate School of Engineering, \\ Nagoya Institute of Technology, Nagoya, Aichi, 466-8555, Japan}
\author{Shinya Sugiura}
\affiliation{Institute of Industrial Science, The University of Tokyo, Tokyo 153-8505, Japan}
\author{Hiroki Wakatsuchi}%
\affiliation{Department of Engineering, Graduate School of Engineering, \\ Nagoya Institute of Technology, Nagoya, Aichi, 466-8555, Japan}

\date{\today}
\begin{abstract}
With their self-tuned time-varying responses, waveform-selective metasurfaces embedded with nonlinear electronics have shown fascinating applications, including distinguishing different electromagnetic waves depending on the pulse width. However, thus far they have only been realized with a spatially homogeneous scattering profile. Here, by modeling a metasurface as time-varying admittance sheets, we provide an analytical calculation method to predict the metasurface time-domain responses. This allows derivation of design specifications in the form of equivalent sheet admittance, which is useful in synthesizing a metasurface with spatiotemporal control, such as to realize a metasurface with prescribed time-dependent diffraction characteristics. As an example, based on the proposed equivalent admittance sheet modeling, we synthesize a waveform-selective Fresnel zone plate with variable focal length depending on the incoming pulse width. The proposed synthesis method of pulse-width-dependent metasurfaces may be extended to designing metasurfaces with more complex spatiotemporal wave manipulation, benefiting applications such as sensing, wireless communications and signal processing.
\end{abstract}

\maketitle

\section{\label{sec:intro} Introduction}
Metamaterials and their two-dimensional versions, metasurfaces, have shown unprecedented methods for wave manipulation, marking a new milestone in the development of functional artificial structures. They are characterized by subwavelength engineered elements, called meta-atoms, and have distinct properties such as a near-zero index \cite{alu2007epsilon}, balanced electric-magnetic responses \cite{pfeiffer2013metamaterial}, or maximal cross-polarization  \cite{ding2015ultrathin}, intended to feature novel functionalities in the macrostructures \cite{yuan2020fully,zhang2021generating}. While many initial works on metasurfaces were on achieving control of a static wave, which does not vary with time, recent developments show dynamic wave shaping via time-varying meta-atoms \cite{ramaccia2019phase,hadad2015space,taravati2019generalized,zhang2019breaking,lImani:2020,zhao2019programmable,hall2021space,liu2018huygens,shaltout2019spatiotemporal,salary2019dynamically,sedeh2020time,zhang2022spatiotemporal,hu2022intelligent}. These active-type metasurfaces have been realized across a wide-range of the electromagnetic spectrum, achieving various interesting spatiotemporal applications such as second harmonic generation, nonlinear beam shaping and non-reciprocity \cite{caloz2019spacetime1,caloz2019spacetime2,taravati2022microwave,tang2019programmable}. However, the realization of these spatiotemporal metasurfaces often requires complicated external excitation subsystems, which could substantially limit the applicability. For example, in controlling individual meta-atom states, complex electrical biasing is required in microwave metasurfaces, which provides advanced control but in return becomes a formidable challenge for a large metasurface area and at higher operating frequencies \cite{li2017electromagnetic,zhao2019programmable,zhang2019breaking}.

To achieve dynamic wave control without a complex chain of external excitation, a resonant metasurface may be combined with nonlinear media. The introduction of nonlinear media transforms the metasurface into a self-tunable device that is sensitive to impinging wave properties such as the power level or polarization state, removing the need for complicated electrical biasing \cite{denz2010nonlinearities,wakatsuchi2013circuit,kiani2020spatial,kiani2020self,quevedo2019roadmap}. In resonant metamaterials, one example of introducing nonlinearity is by inserting nonlinear Kerr nanoparticles within resonant nanorod arrays, which results in the metamaterial having flux-dependent transmission and absorption \cite{chen2010optical}. At microwave frequencies, a method of embedding nonlinear lumped components such as diodes and varactors into resonant metallic meta-atoms has realized various self-tunable power-dependent metamaterials \cite{shadrivov2012metamaterials,luo2015self,luo2019intensity,kim2020active,li2017high}. One of the representative works in Ref.~\cite{shadrivov2012metamaterials} demonstrated microwave beam steering and focusing  depending on the impinging monochromatic light intensity. In recent works, more complex nonlinear power-dependent wave shaping has also been realized, such as obtaining controllable spatial dispersion in a metasurface absorber \cite{luo2022high} or steering surface wave propagation in a mushroom-type metasurface \cite{homma2022anisotropic}. All these works utilized either diodes, varactors, or transistors in realizing nonlinearity within the meta-atom. 

\begin{figure}
    \centering
    \includegraphics[width=\linewidth]{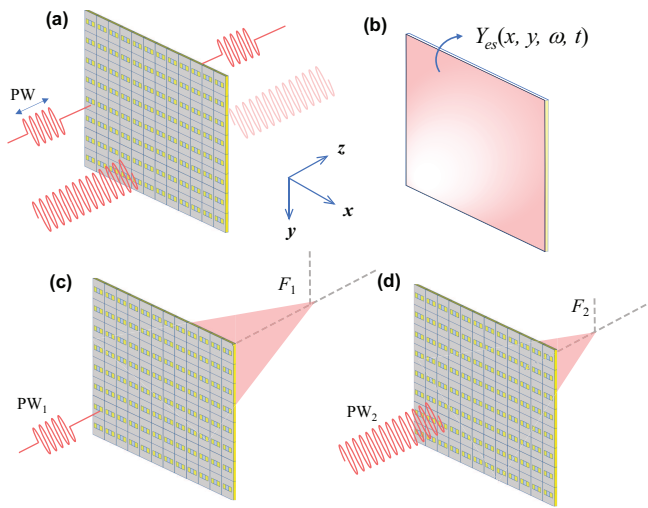}
    \caption{(a) Metasurface with a temporal response enabling waveform selectivity, and (b) its time-varying equivalent admittance. (c, d) Spatiotemporal wave manipulation using the proposed metasurface where the focal length (FL) of the metasurface changes with pulse width (PW) or time. }    \label{fig:1}
\end{figure}

To exploit the temporal scattering of metasurfaces, more recent works have utilized transient responses of reactive circuits under equivalent direct current (DC) input  \cite{wakatsuchi2015waveform,15Wakatsuchi:2013,14Wakatsuchi:2019,ushikoshi2020experimental,kVellucci:2020}. These metasurfaces consist of homogeneous patch structures embedded with a diode bridge full-wave rectifier connected to a combination of a resistor, an inductor, and a capacitor ($R, L, $ and $C$). When an incident microwave beam illuminates the metasurface, the output currents (or voltages) of the diode bridge do not immediately change with time due to the existence of the reactive circuits, producing the effect of time-varying transmission, reflection, and absorption. Unlike other temporal metasurfaces that utilize periodic modulation of meta-atoms to generate sidebands, in these transient metasurfaces, the meta-atoms are designed to transform the incident energy into zero frequency. Their time constant is much longer than the incident beam period; hence, they can perform waveform selectivity or time-domain filtering, i.e., differentiate incident beams depending on how long they propagate through the metasurface even at a constant frequency \cite{wakatsuchi2015time}. An exemplary model of such waveform-selective metasurfaces is shown in Fig.~\ref{fig:1}(a), which allows propagation of an incoming pulse but absorbs or reflects a continuous wave. These metasurfaces can enable various important applications such as designing waveform-selective cloaked antennas \cite{vellucci2019waveform} and realizing reconfigurable surfaces for Wi-Fi signals \cite{ushikoshi2020experimental} as well as for microwave imaging \cite{lImani:2020}. Despite the fascinating properties of these temporal metasurfaces, they have thus far been realized strictly in a spatially homogeneous condition. Moreover, very limited works on analytical modeling of these metasurfaces have been reported, such as in Ref.~\cite{asano2020simplified}, which were focused only on analyzing the transient circuits. Thus, equivalent admittance modeling, commonly used in other thin metasurfaces \cite{holloway2005reflection}, remains unexplored for these transient metasurfaces. 

In this study, we extract the equivalent admittance of a self-tunable temporal metasurface to facilitate design of spatiotemporal wave manipulation. We consider a meta-atom consisting of a subwavelength metallic slit connected to a full-wave rectifier and a combination of $R, L,$ and $C$ components. Such a metasurface can be represented as a self-tuned time-varying electric admittance and a magnetic impedance, in which the time constant corresponds to the circuit parameters being implemented. As seen in Fig.~\ref{fig:1}(b), an equivalent circuit model can be used to accurately predict the admittance $Y_{se}$ and the time-varying scattering parameters at the resonant frequency ($\omega_0$). This analytical modeling provides a convenient design methodology to realize a spatiotemporal metasurface, through which a Fresnel zone plate (FZP) \cite{fowles1989introduction,shams2017740} has been designed that exhibits a time-varying focal point. In this FZP, the focal point changes depending on the pulse width of the incoming wave (Fig.~\ref{fig:1}(c) and \ref{fig:1}(d)). This work represents an important finding on equivalent admittance sheet modeling of temporal metasurfaces and paves the way for subsequent research on self-tuned spatiotemporal metasurfaces. 

\section{Metasurface with a Time-Varying Equivalent Admittance}
\subsection{Metasurface Configuration and Admittance Sheet Modeling \label{sec:ms configuration}}

\begin{figure}[t]
    \centering
    \includegraphics[width=0.96\linewidth]{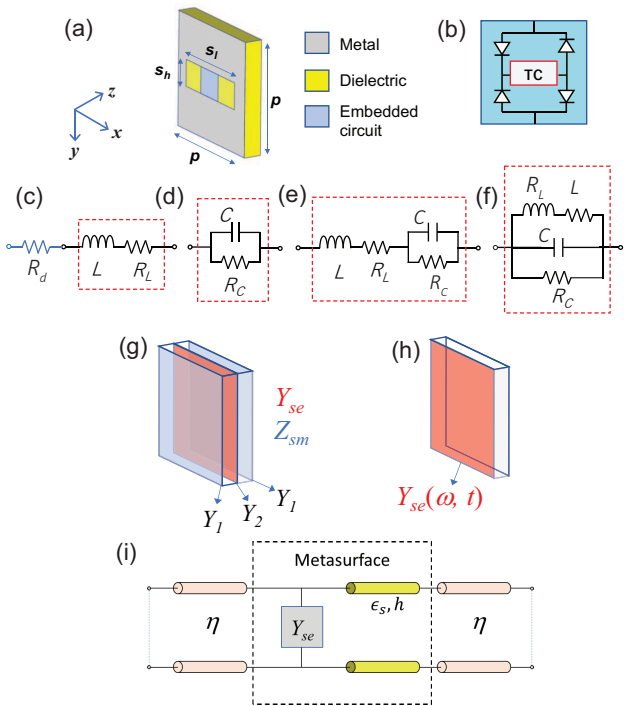}
    \caption{(a) Physical configuration of the self-tuned time-varying metasurface considered within this study. (b) Embedded circuit within the slit: a diode bridge full-wave rectifier loaded with a transient circuit (TC). (c)-(f) Transient circuits with four different transient responses: (c) $RL$ series, (d) $RL$ parallel, (e) $RLC$ series, and (f) $RLC$ parallel circuits. $R_d$ represents the resistive component of two diodes and is connected to all the circuits in series. (g) Ideal model for the metasurface where the scattering properties can be physically represented as three shunt admittances. (h) Metasurface model considered here with only a single-layer shunt admittance $Y_{se}$ due to limited influence of $Z_{sm}$. (i) Equivalent transmission line model of the metasurface with certain dielectric constant $\epsilon_s$ and thickness $h$.}
    \label{fig:2}
\end{figure}

The metasurface considered within this study consists of a thin metallic slit on top of a dielectric layer, as shown in Fig.~\ref{fig:2}(a). The same metasurface has also been reported in Ref.~\cite{14Wakatsuchi:2019}. Within the metallic slit, an embedded circuit is connected. This embedded circuit consists of a diode bridge rectifier loaded with a transient circuit (see Fig.~\ref{fig:2}(b)). The transient circuit can be any combination of $R, L,$ and $C$ components, but for simplification, here, we consider four combinations, i.e., $RL$ series, $RC$ parallel, $RLC$ series and $RLC$ parallel (see Fig.~\ref{fig:2}(c)-(f). Note that in an ideal case, this metasurface unit cell extends into an infinite two-dimensional homogeneous array.

When a transverse electric (TE) wave impinges on the metasurface, it generates a potential difference across the slit. Since the slit is connected to the embedded circuit, when the potential difference across the slit exceeds the diode turn-on voltage, a current starts to flow into the diode bridge and is rectified. With the rectified input voltage or current from the diode bridge, the capacitor or inductor prevents instantaneous changes in both parameters, a response that is commonly found in a DC-driven transient circuit. Therefore, the embedded circuit introduced within the slit is responsible for the transient response observed in the scattering parameters of the metasurface. The temporal scattering profile, e.g., transmission increasing or decreasing with time or increasing then decreasing, depends on the implemented circuit parameters, where the four types of transient circuits in Fig.~\ref{fig:2}(c)-(f) give distinctive results.  

As this metasurface has previously been extensively studied in numerical simulations and experiments \cite{wakatsuchi2015waveform,wakatsuchi2015time,14Wakatsuchi:2019,ushikoshi2020experimental}, here, we focus on  modeling this metasurface as equivalent surface parameters to further assist in the design methodology. In modeling such a transmissive metasurface, since the thickness of the metasurface is much smaller than the wavelength, an impedance boundary condition can be applied, in which the metasurface is equivalent to a distribution of an electric sheet admittance ($Y_{se}$) and a magnetic sheet impedance ($Z_{sm}$). The tangential components of the electric and magnetic field can be equivalent to the voltage and current in circuit theory; hence, the scattering problem of a normally incident plane wave can be modeled by three shunt electric admittances separated by transmission lines, as seen in Fig.~\ref{fig:2}(g). Such modeling has been widely used to describe subwavelength resonant metasurfaces \cite{fathnan2020achromatic,epstein2016huygens,pfeiffer2013metamaterial,monticone2013full,decker2015high}, where, in the case of no magnetoelectric coupling considered, the resonant structures are called Huygens' metasurfaces \cite{epstein2016huygens}. 

Here, since the metasurface only consists of a single-layer metallic structure, the magnetic response is weak. Therefore, in Subsec.~\ref{sec:extraction}, we will show that the modeling of such a metasurface can be realized by considering only a single electric shunt admittance. In this simplified model, the subwavelength slit structure resembles a resonant meta-atom equivalent to a transmission line loaded with a single electric sheet admittance, as shown in Fig.~\ref{fig:2}(h)-(i). 

\subsection{Extraction of the Time-Varying Electric Admittance and Magnetic Impedance \label{sec:extraction}}
Similar to other thin metasurfaces that introduce field discontinuities into the incoming wave, the metasurface considered here can also be treated as a distribution of electric and magnetic sheet impedances \cite{epstein2016huygens,pfeiffer2013metamaterial,monticone2013full,decker2015high}. With the transient response, however, the equivalent admittance and impedance of the metasurface should also have a time dependence. The time-varying sheet impedance and admittance can be obtained by using a calculation with complex transmission and reflection, both having time and frequency dependence, as follows \cite{holloway2005reflection}:
\begin{equation}
\begin{aligned} \label{eq:Yse_1}
 Y_{se}(t,f)=\frac{2(1-T(t,f)-R(t,f))}{\eta(1+T(t,f)+R(t,f))},  
\end{aligned}
\end{equation}
\begin{equation}
\begin{aligned} \label{eq:Zsm_1}
 Z_{sm}(t,f)=\frac{2(1-T(t,f)+R(t,f))}{\eta(1+T(t,f)-R(t,f))}.  
 \end{aligned}
\end{equation}
Here, $T$ is the complex transmission  coefficient ($S_{21}
$), $R$ is the complex reflection coefficient ($S_{11}
$), $f$ is the frequency, $t$ is time, and  $\eta=\sqrt{\mu/\epsilon}$ is the free-space wave impedance. {It is noted that essentially time and frequency are interchangeable and interdependent. The time-frequency representation of  Eqs.~\eqref{eq:Yse_1}-\eqref{eq:Zsm_1} is therefore bounded by a resolution limit where Fourier transform cannot localize simultaneously both the time domain and frequency domain \cite{boashash2015time}. With this limit, a trade-off relation exists in which optimizing frequency resolution leads to poor temporal resolution and vice versa. However, if a system has a slow variation in the time domain, reducing the temporal resolution is acceptable, which in turn would relax the time-frequency resolution bounds. This condition is consistent with the metasurface considered here as it has a much larger time constant $t_c$ than the time period of the waveform $t_w$. Hence, $t_c \gg t_w$ guarantees the validity of Eqs.~\eqref{eq:Yse_1}-\eqref{eq:Zsm_1}.}
\begin{figure}
    \centering
    \includegraphics[width=0.9\linewidth]{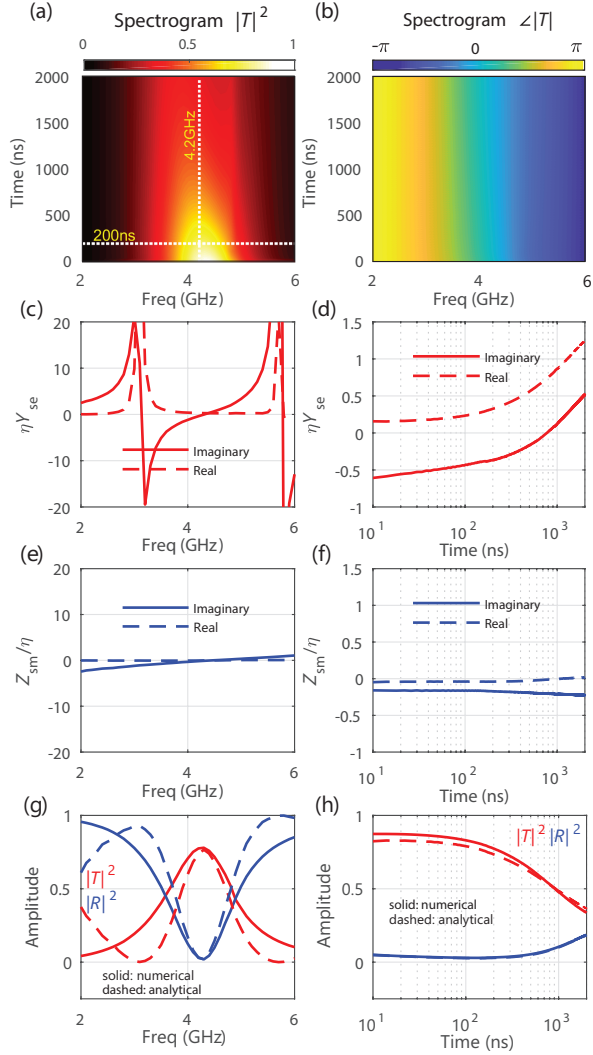}
    \caption{Extraction of equivalent surface parameters of the metasurface. (a) Spectrogram of the metasurface transmission intensity. (b) Spectrogram of the metasurface transmission phase. (c)-(d) $Y_{se}$ and (e)-(f) $Z_{sm}$ (normalized to the wave impedance ($\eta$), both real and imaginary parts) over frequency and time, calculated using Eqs.~\eqref{eq:Yse_1} and \eqref{eq:Zsm_1}. (g) Comparison of transmission intensity over frequency at t=200\,ns between the numerical simulation and analytical calculation using the single shunt admittance model. (h) The same comparison as in (g) over time at 4.2\,GHz. The metasurface here is with $RL$-series circuit where $R_L$=10 $\Omega$ and $L$=0.1 mH. }
    \label{fig:3}
\end{figure}

In extracting the impedance and admittance, we use a co-simulation method based on ANSYS Electronics Desktop 2020 R2 \cite{homma2022anisotropic,wakatsuchi2015waveform,15Wakatsuchi:2013},  as explained in Appendix~\ref{apx:A}. This simulation method integrates a full-wave numerical result within a circuit analysis. The meta-atom, as seen in Fig.~\ref{fig:2}(a), is first analyzed using a full-wave numerical solver, where the meta-atom period is $p=$18\,mm, having slit dimensions of $s_l=$16\,mm and $s_h=$5\,mm, a dielectric thickness of $h=$1.5\,mm and permittivity of $\epsilon_s$=3. In the circuit simulation, the electronic components seen in Fig.~\ref{fig:2}(c)-(f) are embedded, and a transient solver is used in which the results of the circuit simulation are voltages corresponding to the scattering parameters of the meta-atom, i.e., $v_T$ for transmission voltage and $v_R$ for reflection voltage. These voltages are sine waves with certain frequencies but continuously time-varying amplitudes. To be able to calculate Eqs.~\eqref{eq:Yse_1} and \eqref{eq:Zsm_1}, complex representation of the transmission and reflection voltages are needed. Therefore, a postprocessing procedure is implemented for these voltages using the short time Fourier transform (STFT) \cite{sejdic2009time}. 
In the STFT, the voltages are divided into small segments of time before applying the Fourier transform, which can be formulated as
\begin{equation}
{V}_{T,R,I}(\tau_w,f)= \int_{-\infty}^{\infty}   {v}_{T,R,I}(t)w(t-\tau_w)e^{-j2\pi f t} \mathrm{d}t.
\end{equation}
Here, $V_{T}$, $V_{R}$, and $V_{I}$ are Fourier transform outputs from windowed voltages (transmission, reflection, and input voltages, respectively), and $w(t)$ is a window function of length $M$ (4 ns in this study). The window function is centered about time $\tau_w=nP$ where $P$ is the interval time between successive Fourier transforms (set to 2 ns) and $n$ is an integer. 
To obtain consistent phase differences, the transmission phase ($\phi_{T}=\angle{T}$) and reflection phases ($\phi_{R}=\angle{R}$) are calculated by subtracting the phase of a reference input voltage $V_{I}$ from the phases of transmission and reflection voltages, e.g., $\phi_{T}(\tau_w,f)=\angle{{V}_{T}-\angle{V}_{I}}.$
Similarly, the amplitudes of transmission ($A_T=|T|$) and reflection ($A_R=|R|$) are obtained by dividing the amplitudes of the transmission and reflection voltages by that of the reference input voltage $V_{I}$, e.g., $A_T(\tau_w,f)=|V_{T}|/|V_{I}|$. 

Having both time- and frequency-dependent amplitude and phase, we can plot a spectrogram of the metasurface transmission, as seen in Fig.~\ref{fig:3}(a) and (b). Here, we consider a metasurface with an $RL$ series transient circuit (Fig.~\ref{fig:2}(c)), where the transmission is decreasing with time. The inductor here is $L=0.1\,$mH, and the resistor is $R_L=10\,\Omega$. We can see in the transmission amplitude spectrogram of Fig.~\ref{fig:3}(a) that the metasurface is transparent around the resonant frequency of 4.2\,GHz at the initial time, while beyond 1000\,ns, the transmission amplitude is drastically reduced. While this variation in transmission is rapid, the phase variation over time is insignificant, as depicted in Fig.~\ref{fig:3}(b). Using this transmission amplitude and this transmission phase, the complex scattering parameters are then calculated as
\begin{equation}
\begin{aligned}
    T(\tau_w,f)=A_T(\cos{\phi_T}+j\sin{\phi_T}), \\
    R(\tau_w,f)=A_R(\cos{\phi_R}+j\sin{\phi_R}).
\end{aligned}
\end{equation}

Based on the complex $T$ and $R$, we can calculate the electric sheet admittance and magnetic sheet impedance of the metasurface as in Eqs.~\eqref{eq:Yse_1} and \eqref{eq:Zsm_1}, and the result is plotted in Fig.~\ref{fig:3}(c)-(f). Fig.~\ref{fig:3}(c) and (e) shows the electric admittance ($Y_{se}$) and magnetic impedance ($Z_{sm}$) varying with frequency at 200\,ns, as indicated by the horizontal dashed line in Fig.~\ref{fig:3}(a). Fig.~\ref{fig:3}(d) and (f) shows the electric admittance ($Y_{se}$) and magnetic impedance ($Z_{sm}$) varying with time at 4.2\,GHz, as indicated by the vertical dashed line in Fig.~\ref{fig:3}(a). 

We see in Fig.~\ref{fig:3}(e) and (f) a modest variation in the magnetic impedance $Z_{sm}$ with time and frequency. This is in contrast to the electric admittance $Y_{se}$ in Fig.~\ref{fig:3}(c) and (d), where the variation with time and frequency is significant. Therefore, the magnetic contribution to the overall scattering properties is expected to be minimal. This corresponds to the fact that the metasurface here consists of only a single metallic structure supported by a thin dielectric. Although the entire physical representation of such a transmissive metasurface contains three shunt electric admittances separated by dielectric layers, as seen in Fig.~\ref{fig:2}(g), here, we simplify the physical model into a single shunt electric admittance, as seen in Fig.~\ref{fig:2}(h)-(i). In this simplified model, we only consider the electric sheet admittance of the meta-atom $Y_{se}$ as the main contributor to the scattering properties. In the two-port network representation, based on the $ABCD$ matrix, we can analytically calculate the scattering profile of the metasurface (see Appendix~\ref{apx:B}). The result is depicted by the dashed lines in Fig.~\ref{fig:3}(g)-(h), where we see that the calculated scattering parameters agree well with the numerical simulation, especially near the resonant frequency of 4.2\,GHz. At lower and higher frequency regions, relatively large discrepancies appear as magnetic impedance sheets ($Z_{sm}$) are omitted for the sake of simplicity. However, this result ensures that the single equivalent admittance is adequate for modeling the temporal response of the metasurface at the resonant frequency (Fig.~\ref{fig:3}(h)).

\subsection{Equivalent Circuit Model \label{sec:equivalent_circuit}}
To calculate the temporal response of the metasurface and assist in the design methodology, we use an equivalent circuit model for the shunt admittance $Y_{se}$. This equivalent circuit model is important to enable the design of spatiotemporal metasurfaces as it {provides} a direct relation between a meta-atom temporal response and its embedded circuit parameters. 

{As the metasurface here exhibits slow time variations relative to the time period of the waveform, the time-frequency resolution limit is relaxed (as discussed after Eqs.~{\eqref{eq:Yse_1}}- \eqref{eq:Zsm_1}). Therefore, two independent physical contributions can be used to describe the scattering properties:} first, a shunt admittance that varies with frequency $Y_{freq}(f)$, which originates from the dispersion characteristics of the metallic slit, and second, a shunt admittance that varies with time $Y_{time}(t)$, which originates from the embedded circuit transient responses. Note that $Y_{freq}(f)$ may contain extra parameters that relate to the frequency dependence, for instance, the parasitic capacitance of diodes. {Also, with the relaxed time-frequency resolution limit, both $Y_{freq}(f)$ and $Y_{time}(t)$ appear independent from each other. Therefore, we can take their summation as the total shunt admittance}, as follows:
\begin{equation} \label{eq:Yse_tf}
    Y_{se}(t,f)=Y_{freq}(f)+Y_{time}(t).
\end{equation}
For a normally incident plane wave, the metallic slit structure resembles a resonant element that has dispersion characteristics similar to a parallel $LC$ circuit \cite{fathnan2020bandwidth}. Hence, the frequency-dependent admittance $Y_{freq}(f)$ from Eq.~\eqref{eq:Yse_tf} can be attributed to a simple $LC$ admittance as follows:
\begin{equation} \label{eq:Yse_f}
    Y_{freq}(f)=j\bigg(\omega C_m-\frac{1}{\omega L_m}\bigg)+\dot{a},
\end{equation}
\begin{equation}
    \dot{a}=a'+ja'',
\end{equation}
where $L_m$ and $C_m$ are the circuit parameters for the metallic slit, which can be extracted from a full-wave simulation \cite{olk2019accurate,fathnan2020bandwidth}, and $\omega$ is the angular frequency. In the above equation, a complex adjustment parameter $\dot{a}$ is added to account for a resonance shift that may be observed in a numerical simulation or experiment. 

\begin{figure}
    \centering
    \includegraphics[width=\linewidth]{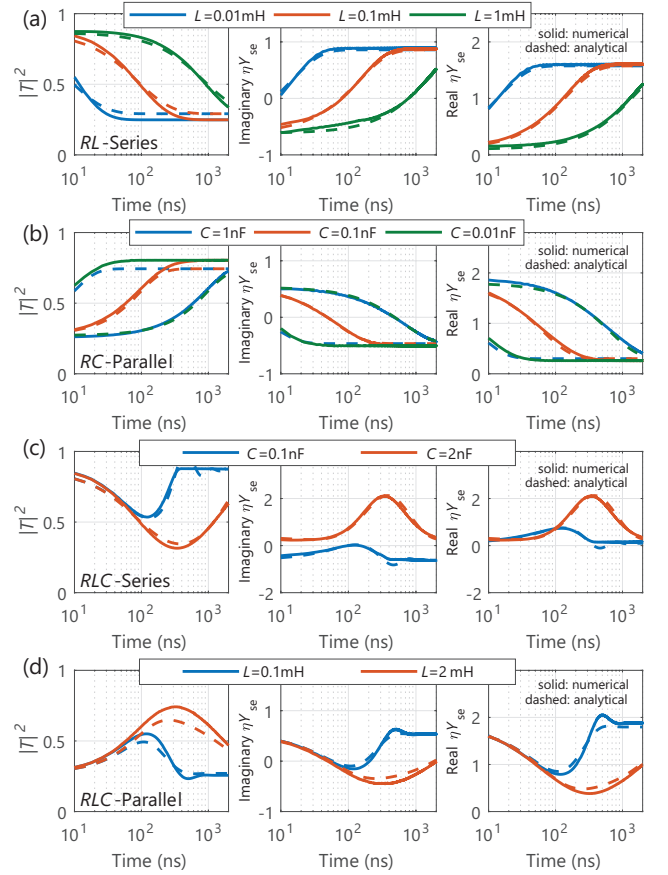}
    \caption{Analytically calculated complex sheet admittances and corresponding transmission intensity (dashed lines) compared to the numerical simulation (solid lines) (a) for an $RL$ series metasurface with varied $L$ and fixed $R_L$=10\,$\Omega$, (b) for an $RC$ parallel metasurface with varied $C$ and fixed $R_C$=10\,k$\Omega$, (c) for an $RLC$ series with varied $C$ and fixed $L$=0.1\,mH, $R_L$=10\,$\Omega$, and $R_C$=10\,k$\Omega$, and (d) for an RLC parallel with varied $L$ and fixed $C$=0.1\,nF, $R_L$=10\,$\Omega$, and $R_C$=10\,k$\Omega$. In these analytical calculations, complex adjustment parameters $\dot{a}$ and $\dot{b}$ presented in Table~\ref{tab:adjustment_factors} are used.}
    \label{fig:4}
\end{figure}

The second part of the admittance in Eq.~\eqref{eq:Yse_tf}, which is a time-dependent admittance $Y_{time}(t)$, emerges mainly due to the transient response of the embedded circuit. Hence, it can be quantified based on the time evolution of currents or voltages within the embedded circuit. In this equivalent circuit modeling, we assume that the incident wave is fully rectified and that the circulating current is DC. In such a DC-driven circuit,  we have a transient current $i(t)$ and an initial applied potential $E_0$; therefore, we can have $i(t)/E_0$ as a DC admittance. Since this DC admittance is directly related to the current induced by the incident wave, it should also depend on the coupling between the incident wave and the metasurface or the embedded circuit system. Therefore, to adjust the significance of the coupling, we assign another complex parameter $\dot{b}$, yielding
\begin{equation} \label{eq:Yse_t}
    Y_{time}(t)=\dot{b}\frac{i(t)}{E_0},
\end{equation}
 \begin{equation}
     \dot{b}=b'+jb''.
 \end{equation}

As the DC admittance $i(t)/E_0$ in Eq.~\eqref{eq:Yse_t} is dependent on the $RLC$ circuit configuration, it is predictable by an equivalent circuit model. In Appendix~\ref{apx:C}, we present a derivation of the DC admittance for the four circuit types seen in Fig.~\ref{fig:2}(c)-(f), as has also been partially reported in Ref.~\cite{asano2020simplified}. Particularly, analytical expressions of the DC admittances for the four circuit types are formulated in Eqs.~\eqref{eq:Ydc_L}, \eqref{eq:Ydc_C}, \eqref{eq:Ydc_series}, and \eqref{eq:Ydc_parallel}, and have been validated by a circuit simulation as detailed in Appendix~\ref{apx:D}. Throughout this equivalent circuit modeling, the diode bridge is modeled as resistor $R_d$ connected in series with the transient circuit (see Fig.~\ref{fig:2}(c)). {Note that since in the rectification process electric charges pass two diodes in series connection, therefore $R_d$ is double the effective resistance of a single diode $R_0$, i.e., $R_d=2 R_0 (= 680\,\Omega$ in this study) \cite{asano2020simplified}.} The same series $R_d$ connection should also be present for the other transient circuits in Fig.~\ref{fig:2}(d)-(f), although we only draw it in Fig.~\ref{fig:2}(c). 
In calculating the time-varying admittance, both $\dot{a}$ and $\dot{b}$ in Eqs.~\eqref{eq:Yse_f} and \eqref{eq:Yse_t} are chosen by fitting to a numerical or experimental result (see the last paragraph of Appendix~\ref{apx:C} for details). Analytically calculated time varying admittances from this circuit modeling are shown by dashed lines in Fig.~\ref{fig:4} for both the real part (right column) and imaginary part (middle column). The corresponding calculated transmission is obtained by the transmission line model (Appendix~\ref{apx:B}). Four different transient circuits, (a) $RL$ series, (b) $RC$ parallel, (c) $RLC$ series, and (d) $RLC$ parallel are considered. In these four cases, the inductance and capacitance are varied to observe the practicality of the circuit modeling.
For a comparison, we numerically simulate the four metasurface types using the co-simulation method and extract the complex admittances using Eq.~\eqref{eq:Yse_1}.  As depicted in Fig.~\ref{fig:4}, the numerical simulations (solid lines) and analytical calculations (dashed lines) agree with each other. This result indicates that the circuit modeling efficiently represents the meta-atom responses and is useful in designing a metasurface with variation in temporal scattering properties.  

\subsection{Full-wave Simulation of the Time-Varying Admittance \label{sec:full-wave MA admittance}}
\begin{figure}
    \centering
    \includegraphics[width=0.9\linewidth]{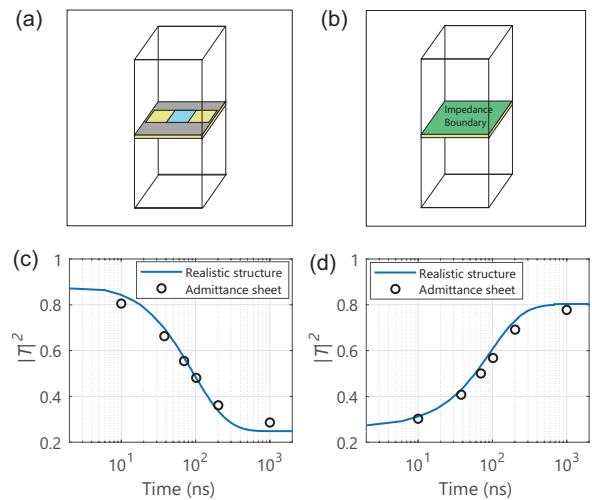}
    \caption{Comparison between the metasurface simulation using the realistic structure and the admittance sheet model at 4.2 GHz over time. (a) Simulation setup as shown in ANSYS Electronics Desktop 2020 R2 using the realistic structure (co-simulation) and (b) using admittance sheet model. (c) Transmission intensity for the metasurface with the $RL$ series transient circuit having $L$=0.1 mH and $R_L$=10 $\Omega$. (d) Transmission intensity for the metasurface with the $RC$ parallel transient circuit having $C$=0.1 nF and $R_C$=10 k$\Omega$.}
    \label{fig:5}
\end{figure}

As the temporal admittance of the metasurface can be extracted and analytically predicted, a full-wave simulation based on the admittance sheet model may be conducted to obtain the scattered field distribution from the metasurface at a particular time. In this admittance model, we use a dielectric slab and one impedance boundary condition. The simulation setup using ANSYS Electronics Desktop 2020 R2 is shown in Fig.~\ref{fig:5}(a) and (b), comparing the physically realistic model based on the co-simulation method (a) with the simplified admittance sheet model (b). In the admittance sheet model, a separate simulation is performed for different time samples. For example, here, we simulate the admittance profile in six samples at 10\,ns, 34\,ns, 70\,ns, 100\,ns, 200\,ns, and 1000\,ns. In each simulation, the scattering parameters and the scattered fields can be directly obtained without a post-processing procedure.

Simulation results for the metasurfaces with $RL$ series and $RC$ parallel transient circuits are shown in Fig.~\ref{fig:5}(c) and (d), respectively. For the $RL$ series case, the inductor is $L$=0.1\,nH, and the resistor is $R_L$=10\,$\Omega$, while for the $RC$ parallel case, the capacitor is $C$=0.1\,nF, and the resistor is $R_C$=10\,k$\Omega$. The admittance sheet model (circles) and the physically realistic simulation model (blue line) results agree with each other. A slight discrepancy can be observed due to the admittance sheet model not taking into account the magnetic response $Z_{sm}$, which, as discussed in Sec.~\ref{sec:ms configuration}, has very little influence on the overall scattering parameters.

\begin{SCfigure*}
\centering
    \includegraphics[width=1.6\linewidth]{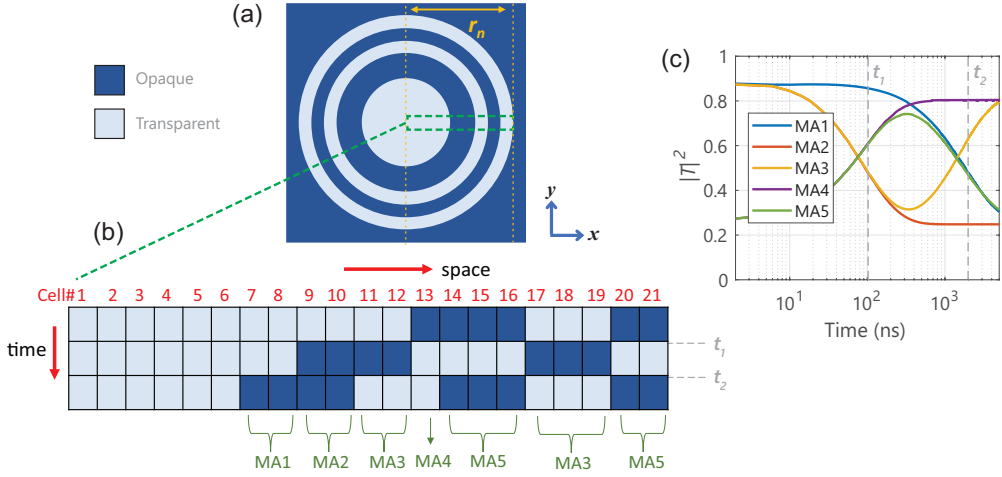}
    \caption{(a)-(b) Design of a binary amplitude self-tuned FZP with the distribution of the meta-atom transmission profile varying in space and time. The meta-atom size is 18\,mm, and the designed focal distance is $F$=500, 300, and 200 mm at three different times. The diameter of the FZP is $d$=756\,mm. (c) Simulated temporal transmission intensity profile of five meta-atoms used for the time-varying FZP.}
    \label{fig:6}
\end{SCfigure*}

\section{Application to Self-Tuned Spatiotemporal Wave Manipulation \label{sec:III}}
\subsection{Design of a Fresnel Zone Plate with Time-Varying Focal Lengths}
In the previous subsection, we have modeled the metasurface as a time-varying admittance and confirmed that a full-wave simulation of the admittance sheet model agrees with the physically realistic metasurface simulation. Now, we use the method to synthesize a metasurface with a more complex spatiotemporal configuration. Since the metasurface discussed here has temporal variation of the transmission amplitude while exhibiting a relatively constant phase, we can utilize it to realize prescribed time-varying diffraction characteristics. As an example, here, we synthesize a Fresnel Zone Plate (FZP) with a variable focal point depending on the incoming pulse width, as illustrated in Fig.~\ref{fig:1}(c)-(d).

In FZPs, opaque and transparent regions within concentric rings alternate over a two-dimensional surface to obtain constructive interference at a focal distance $F$. {Only the transmission amplitude varies in these regions while the transmission phase should be the same}. As seen in Fig.~\ref{fig:6}(a), the ring radii have a binary transmission amplitude (dark blue for opaque shields and light blue for transparent windows), and the radius $r_n$ changes by zone number $n$. The key to realizing a time-varying FZP is to engineer the distribution of the binary transmission amplitude over time such that the focal point is prescribed at different distances for different times, or equivalently pulse widths. Using Fresnel zone theory \cite{fowles1989introduction}, the radii for the time-varying FZP can be formulated as
\begin{equation} \label{eq:FZP-formula}
    r_n(t)=\sqrt{n\lambda F(t)+\frac{1}{4}n^2\lambda^2}.
\end{equation}
Here, $\lambda$ is the operating wavelength. Note that the focal distance $F$ has a prescribed variation with time $t$.

To realize this time-varying focal point FZP, we consider the meta-atom in Sec.~\ref{sec:ms configuration} with four different transient circuits. Using these meta-atoms as building block of the FZP, we may obtain a time-varying distribution of opaque and transparent regions, {while maintaining almost the same phase}. As the FZP requires a binary amplitude profile, using the proposed metasurfaces, in accordance with Eq.~\eqref{eq:FZP-formula}, three different focal points can be designed for three time segments. Fig.~\ref{fig:6}(b) depicts the distribution of opaque and transparent meta-atoms over space and time in half of the FZP diameter ($d/2$).  Note that to simplify the design process, here we consider only one-dimensional (1D) profile of the FZP along the horizontal axis. Similar to the metasurface described in Sec.~\ref{sec:extraction}, the size of each meta-atom is 18\,mm, and there are 21 cells arranged in half of the diameter ($d=756$\,mm) of the FZP, as indicated by cell numbers 1-21. The operating frequency is 4.2\,GHz ($\lambda$=71\,mm). The dashed square in Fig.~\ref{fig:6}(a) indicates the particular zoomed-in space in Fig.~\ref{fig:6}(b). The FZP is designed to have a decreasing focal point with time, where at an initial time range $t<t_1=$100\,ns, the focal distance is $F=500\,$mm ($F/d=0.66$), at an intermediate time range $t_1<t<t_2=$2000\,ns, the focal distance is $F=300$\,mm ($F/d=0.4$), and in the steady-state condition at $t>t_2$, the focal distance is $F=200\,$mm ($F/d=0.27$). As shown in Fig.~\ref{fig:6}(b), if we consider the time domain and that each cell represents a meta-atom with a certain time-varying transmission, then we obtain five different meta-atom types (MA1-MA5). From cells 1 to 6, the transmission is always the maximum; hence, they can be left blank, i.e., only a dielectric without metal patterns or circuits. For cells 7-21, the transmission profile changes over time and can be represented by different meta-atom designs. For example, in cells 7 and 8, the transmission is always the maximum before $t_2$. This time-varying transmission profile is equivalent to a meta-atom with the $RL$ series transient circuit. Similarly, for cells 14, 15 and 16, the transmission is the maximum only in the intermediate time of $t_1<t<t_2$, which is equivalent to the transmission of a meta-atom with the $RLC$ parallel transient circuit. 

With the required spatiotemporal distribution of opaque and transparent regions for the FZP, we design meta-atoms suitable for the realization. Accordingly, in each meta-atom, we choose circuit parameters to satisfy the binary amplitude profile in Fig.~\ref{fig:6}(b). As the transmission of the metasurface is directly related to the exponential terms of the corresponding DC admittance, we can reduce the problem of finding suitable circuit parameters by only considering the exponential decay constants (see exponential terms in Eqs.~\eqref{eq:Ydc_L}, \eqref{eq:Ydc_C}, \eqref{eq:Ydc_series}, and \eqref{eq:Ydc_parallel} in Appendix~\ref{apx:C}). First, we assume that the boundary between transparent and opaque regions is 50\% transmission (e.g., transparent if $|T|>$0.5). Therefore, we can use the exponential decay terms to find the suitable $L$ or $C$ to realize the transmission profile at a particular designed time. For the metasurface with an $RL$ series circuit, if we take the exponential term in Eq.~\eqref{eq:Ydc_L} and set the exponential decay to 50\%, i.e., $e^{\left(-\frac{R_L+R_d}{L}\right)t}=0.5$, then we obtain
\begin{equation} \label{eq:L_tb} 
    L=-t_b\frac{R_LR_d}{\mathrm{ln}0.5}.
\end{equation}
Similarly, for the metasurface with an $RC$ parallel circuit, using the exponential term in Eq.~\eqref{eq:Ydc_C}, by setting $e^{-\frac{RC+R_d}{CR_CR_d}t}=0.5$, we obtain
\begin{equation} \label{eq:C_tb}
    C=-t_b\frac{R_C+R_d}{(\mathrm{ln}0.5)R_CR_d}.
\end{equation}
Here, $t_b$ is the time in which the exponential decay reaches 50\% of its initial value. Using these equations, we calculate the required $L$ and $C$ for MA1 to MA5 by setting $t_1$=100\,ns and $t_2$=2000\,ns. For example, in MA1, as seen in Fig.~\ref{fig:6}(b), the boundary between opaque and transparent regions occurs at $t_2$. Thus, by substituting $t_b=2000$\,ns and $R_L$=10\,$\Omega$ into Eq.~\eqref{eq:L_tb}, we obtain $L$=0.1\,mH. For MA2, using the same method, we obtain $L$=2\,mH. For the $RLC$ series and parallel cases, the corresponding $L$ and $C$ within the circuit can be calculated using the same equations, as these cases are a combination of the $RL$ series and the $RC$ parallel cases. In the $RLC$-series case, Eq.~\eqref{eq:L_tb} is used to obtain $L$ when the transmission decreases from the maximum to 50\% at $t_1$, while Eq.~\eqref{eq:C_tb} is used to obtain $C$ when the transmission increases from the minimum to 50\% at $t_2$. A similar calculation is performed for the $RLC$ parallel case, but with an inverse temporal transmission profile. In Fig.~\ref{fig:6}(c), the transmissions of designed meta-atoms MA1 to MA5 are plotted. Details of the meta-atom types used and the circuit parameters are presented in Table~\ref{tab:MA_parameters}.

\begin{figure*}
    \centering
    \includegraphics[width=0.82\linewidth]{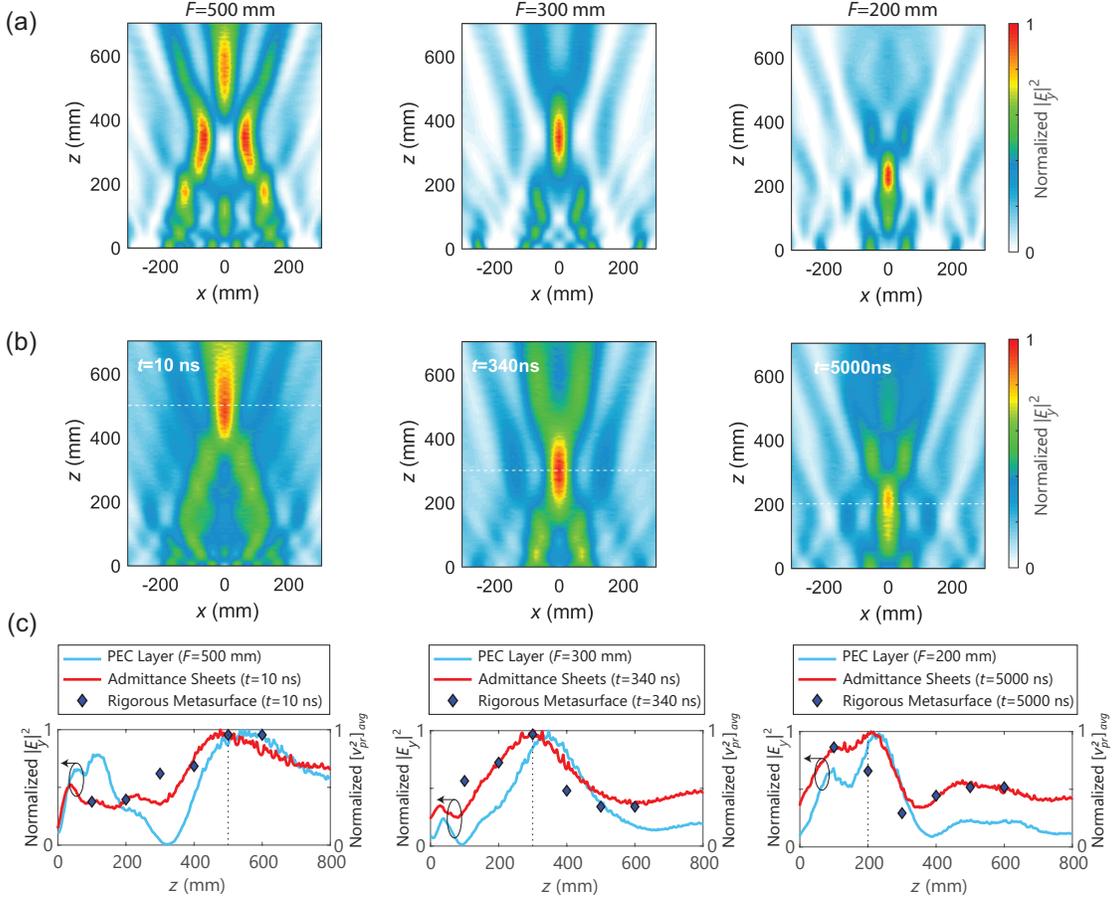}
    \caption{Binary amplitude FZP with a pulse-width-dependent focal point. Simulated electric field intensity (normalized ${|E_y|}^2$) from (a) ideal static binary FZPs based on PEC layers with different focal point designs and (b) FZPs based on equivalent admittance sheets that represent three fixed time points. (c) The electric field intensity at $x$=0 mm (normalized ${|E_y|}^2$) for the FZPs based on the PEC layers and the equivalent admittance sheets (red and blue curves, left $y$-axes) compared to the probe voltage (normalized $[v_{pr}^2]_{avg}$) for full FZP based on the rigorous waveform-selective metasurface (diamonds, right $y$-axes).} 
    \label{fig:7}
\end{figure*}

\begin{table}[h]
\caption{Meta-atom circuit parameters for the binary amplitude self-tuned FZP \label{tab:MA_parameters}}
\begin{ruledtabular}
\begin{tabular}{lllllll}
    & Circuit type & \textit{$R_d$} & \textit{$R_L$} & \textit{$R_C$} & \textit{L}     & \textit{C}    \\ \hline
MA1 & \textit{RL} Series & 680\,$\Omega$ & 10\,$\Omega$ &     - & 2\,mH   & -     \\ \hline
MA2 & \textit{RL} Series & 680\,$\Omega$ & 10\,$\Omega$ &     -     & 0.1\,mH & -     \\ \hline
MA3 & \textit{RLC} Series & 680\,$\Omega$ & 10\,$\Omega$ &     10\,k$\Omega$   & 0.1\,mH & 1\,nF   \\ \hline
MA4 & \textit{RC} Parallel & 680\,$\Omega$ & - &     10\,k$\Omega$     & -     & 0.1\,nF \\ \hline
MA5 & \textit{RLC} Parallel & 680\,$\Omega$ & 10\,$\Omega$ &     10\,k$\Omega$ & 2\,mH   & 0.1\,nF \\ 
\end{tabular}
\end{ruledtabular}
\end{table}

\subsection{Numerical Validation of the Spatiotemporal Response}

To confirm the designed FZP with a time-varying focal point, we first simulate an ideal static FZP case. In this simulation, the same transparent and opaque region as illustrated in Fig.~\ref{fig:6}(b) is used where only perfect electric conductor (PEC) layers on top of the dielectric substrate are simulated (see Appendix~\ref{apx:E} for details). Three simulation instances are conducted, and the resulting $y$-polarized electric field intensity are shown in Fig.~\ref{fig:7}(a), where three different focal points can be seen located at approximately 500, 300, and 200\,mm. Note that since the FZPs are discretized with a cell size of 18\,mm, the ideal Fresnel zone radius $r_n$ may not be accurately realized for all the instances. As a result, a slight shift in the focal distance $F$ from the designed location may occur as seen from Fig.~\ref{fig:7}(a). After simulating this ideal static FZP case, we simulate the time-varying vocal point of the FZP; first using the admittance sheet model and then using a rigorous waveform-selective metasurface based on the co-simulation method. In the admittance sheet model, the method explained in Subsec.~\ref{sec:full-wave MA admittance} is used, in which the meta-atoms are simulated at three specific times, i.e., $t$=10\,ns, 340\,ns, and 5000\,ns. At each time, the admittance distribution corresponding to MA1 to MA5 is utilized to obtain the $y$-polarized electric fields scattered in front of the FZP. The result is presented in Fig.~\ref{fig:7}(b), where we can see that focal point changes at the three different time instances, corresponding to the self-tuned admittance of the designed metasurface.

To further validate the time-varying focal points, we simulate the full FZP including the rigorous waveform-selective metasurface by using the co-simulation method inside ANSYS Electronics Desktop 2020 R2. Since it is not possible to obtain the transient scattered field in the co-simulation method, a monopole antenna is used as a probe to capture the time-varying voltages within the scattering area (see Appendix~\ref{apx:E} for details). In contrast to the previous simulation (the FZP based on admittance sheets) which gives scattered field intensity profiles for specific discretized instants, this simulation enables us to produce voltages varying in the time domain at the location of the probe. 
Fig.~\ref{fig:7}(c) depicts one-dimensional field profiles of the FZP based on the rigorous waveform-selective metasurface (diamonds) as well as those based on the ideal PEC layers (blue curve) and the equivalent admittance sheets (red curve). We see that despite a slight shift in the maximum level of voltages in the steady state condition ($t$=5000\,ns in Fig.~\ref{fig:7}(c)), the time-varying focal point is also confirmed by the full FZP using the rigorous waveform-selective metasurface. This result supports that the metasurface-based FZP exhibits the desired spatiotemporal response, i.e., the time-varying focal point, and that the design methodology proposed here is valid in realizing such a spatiotemporal metasurface. {The focusing efficiencies calculated as a ratio between the field intensities in the focal plane with and without the presence of the metasurface are 43.59\,\%, 41.74\,\%, and 39.58\,\%, for \textit{t}\,=\,10\,ns, 340\,ns and 5000\,ns, respectively. These efficiencies are lower than other impedance-matched metasurfaces \cite{fathnan2020achromatic} due in part to the losses considered in the admittance model as well as reflected waves in the opaque zones of the FZP.}

\section{Discussion}

It should be noted that the circuit modeling presented here accurately predicts the temporal response of the metasurface only near the resonant frequency. In a lower/higher frequency range, the meta-atom model should be modified to include not only the electric responses $Y_{se}$ but also the magnetic responses $Z_{sm}$. In addition, the presented equivalent circuit model requires prior information of a numerical or experimental result for an accurate prediction of the time-varying admittance. Therefore, future studies may be aimed in obtaining analytical formulations for both the coupling of the metasurface to the incident wave and frequency shifts due to parasitic elements. While the extraction of time-varying admittance is applied to only a transmissive meta-atom structure here, the method can also be implemented to a reflective meta-atom structure \cite{14Wakatsuchi:2019}. {The time-varying admittance concept discussed in Subsec.~\ref{sec:extraction} is applicable to various frequency bands, although the corresponding analytical model in Subsec.~\ref{sec:full-wave MA admittance} may require adjustments in the optical bands due to higher ohmic losses involved in most metallic nanostructures}.  

Despite the fact that our study only shows an FZP with a predesigned time-varying focal point, various other spatiotemporal metasurfaces could be envisaged based on the proposed design and simulation method. For example, the binary FZP has been widely used as a beam-steering device when a spherical source is placed at its focal point with variable lateral shift \cite{shams2017740}. Similar time-space variation for such an FZP with a spherical source may also be conducted in which the time-varying admittance mimics the lateral shift of the spherical source. Such a metasurface can perform beam-steering based on pulse width of the incoming wave. Moreover, in the present work, we only consider a metallic slit structure embedded with nonlinear circuits, which gives weak magnetic responses. Therefore, almost no time-varying phases was observed from the scattering properties. Further improvements can be achieved by designing a metasurface with both time-varying electric and magnetic sheet impedances, which could lead to time-varying phase responses.

With such innovative engineering and tunability, the proposed metasurface may find a variety of applications in communications and signal processing. For example, recently, metasurface-based intelligent reflecting surfaces (IRSs) have been extensively investigated for the next-generation wireless standard \cite{wu2019intelligent,ozdogan2019intelligent,sugiura2021joint}. Most existing studies assume that each unit cell of IRSs is connected with a control line, and the input voltage is optimized to form a reflected beam toward a targeted direction \cite{sugiura2021joint,dai2020reconfigurable,zhang2018space}. However, such a configuration with multiple control lines increases the hardware cost, especially when the number of IRS elements is huge. Furthermore, it is typically assumed that the IRSs are precisely synchronized with a base station in a symbol or frame level, which also escalates the complexity of the system. By contrast, the IRS based on waveform-selective metasurfaces is capable of combating the above-mentioned limitations as the explicit benefits of our pulse width-dependent metasurface tuning. More specifically,  the beam of the proposed metasurface is controlled by optimizing the pulse width, which does not need any control lines connected to the unit cells or precise synchronization between the base station and the IRSs.

\section{Conclusion}
Analytical and numerical studies of a self-tuned time-varying metasurface have been conducted, considering a metasurface with a metallic slit structure embedded with nonlinear and transient electronic circuits. With the metasurface having a low profile and deeply subwavelength dimensions, it can be modeled as a single electric shunt admittance connected to a transmission line. Since the time constant of the transient variation is much larger than the time period of the waveform, we decompose the admittance into two independent parameters to account for both the metallic slit frequency dispersion and the embedded circuit time dependence. We show that such time- and frequency-dependent sheet admittances can reconstruct the scattering profile of the metasurface, where the analytical calculation results agree well with the numerical simulation results. Four different transient circuits are used in this study, each corresponding to a distinct temporal scattering profile.

We then confirm that such a time-varying admittance can be simulated using an impedance boundary condition layer within a commercially available full-wave simulator (ANSYS Electronics Desktop HFSS). The scattering parameters of the metasurface at a particular time simulated by the admittance sheet model are consistent with the results obtained using the physically realistic structure. Therefore, the full-wave simulation method based on the admittance sheet can be used to predict the temporal response from a more complicated structure, such as from an inhomogeneous metasurface. This can complement conventional co-simulation methods, which do not provide time-dependent scattered fields for such time-varying surfaces. As an exemplary model, here, we use the proposed concept to design an FZP where the focal point changes depending on time or the pulse width of the incoming wave. The presented metasurface holds great potential to achieve other self-tunable time-varying wave control, and the admittance sheet modeling offers a versatile method to design such spatiotemporal metasurfaces. 

\begin{acknowledgments}
This work was supported in part by the Japanese Ministry of Internal Affairs and Communications (MIC) under the Strategic Information and Communications R\&D Promotion Program (SCOPE) No. 192106007, the Japan Science and Technology Agency (JST) under the Precursory Research for Embryonic Science and Technology (PRESTO) No. JPMJPR1933 and JPMJPR193A, and the Japan Society for the Promotion of Science (JSPS) KAKENHI No. 17KK0114 and 21H01324.
\end{acknowledgments}

\appendix

\section{Metasurface Co-simulation Method \label{apx:A}}
\begin{figure}[t]
    \centering
    \includegraphics[width=\linewidth]{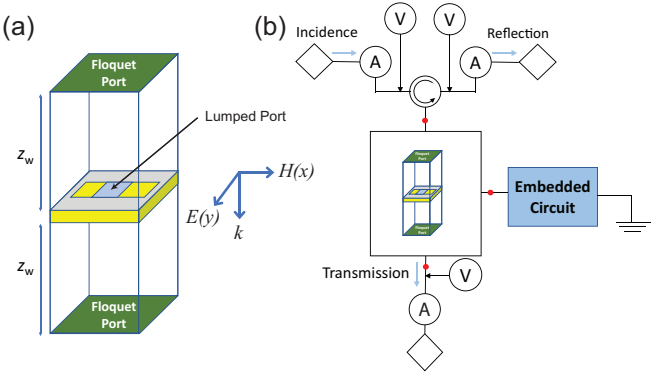}
    \caption{Co-simulation method integrating electromagnetic simulator with circuit simulator. (a) Schematic of the 3D full-wave simulation of the metasurface using periodic boundaries and Floquet ports. (b) Schematic of the circuit simulation. The metasurface full-wave simulation result is imported into the circuit simulator as an embedded circuit component, along with other necessary components connected to it.}
    \label{fig:apx1}
\end{figure}

ANSYS Electronics Desktop 2020 R2 is used to simulate the metasurface in Fig.~\ref{fig:2}(a), based on a co-simulation method  that integrates a full-wave numerical simulation with a circuit simulation. In such a simulation method, the electromagnetic field distribution cannot be obtained since the final results are produced by circuit simulations. A 3D full-wave simulation of a single meta-atom is first conducted using periodic boundaries and Floquet port excitation, as depicted in Fig.~\ref{fig:apx1}(a). The distances between the metasurface and the Floquet ports are $z_w=100$\,mm. Within the metallic slit, a 5\,mm $\times$ 5\,mm lumped port is connected to represent a connection of the embedded circuit. Next, we import a model based on the full-wave simulation result of the meta-atom into a circuit simulation, as depicted in Fig.~\ref{fig:apx1}(b). In the circuit simulation, the embedded circuit consisting of a diode bridge rectifier and $RLC$ components (see Fig.~\ref{fig:2}~(a)-(f)) are connected. Other necessary components are also placed within the circuit simulation to obtain time-varying currents and voltages at the corresponding transmission and reflection ports. For the diodes, we use the SPICE model based on {Avago's HSMS 286x} Schottky diodes series as detailed in Table~\ref{tab:spice_model}. The input power in the circuit simulation is set to 0\,dBm and a transient solver is used within a time span of 10\,$\mu$s and over a frequency variation range from 2 to 6\,GHz with a 0.1\,GHz step. Using this simulation setup, we can monitor the voltages in the Floquet ports (both at the transmission and reflection sides), which has both time and frequency dependency. 

\begin{table}[h]
\caption{SPICE parameters for the diodes \label{tab:spice_model}}
\begin{ruledtabular}
\begin{tabular}{llll}
SPICE parameter & Value    & SPICE parameter & Value    \\ \hline
$B_V$          & 0.7\,V     & \textit{N}               & 1.08     \\
$C_{J0}$         & 0.18\,pF  & $R_S$              & 6.0\,$\Omega$    \\
\textit{EG}      & 0.69\,eV  & $P_B (VJ)$         & 0.65\,V   \\
$I_{BV}$         & 1$\times$10$^{-5}$\,A & $P_T (XTI)$        & 2        \\
$I_S$          & 5$\times$10$^{-8}$\,A & \textit{M}               & 0.5     \\
\end{tabular}
\end{ruledtabular}
\end{table}

\section{Two-Port Network Representation \label{apx:B}}
To analytically obtain the scattering parameters of the metasurface, we use a two-port network representation. In this method, $ABCD$ transfer matrices of the sheet admittance \textit{\textbf{M$_{se}$}} and the transmission line \textit{\textbf{M$_{tl}$}} are utilized to obtain an $ABCD$ transfer matrix of the metasurface \textit{\textbf{M$_{ms}$}}, formulated as

\begin{equation} \label{eq:m_ms}
\bm{M_{ms}}=\bm{M_{se}}\cdot\bm{M_{tl}},
\end{equation}
with 
\begin{align}
    \bm{M_{se}} =
\begin{bmatrix}
1 & 0\\
Y_{se} & 1
\end{bmatrix},
\end{align}

\begin{align}
    \bm{M_{tl}} =
\begin{bmatrix}
\cos(\beta_s h) & jZ_{s}\sin(\beta_s h)\\
jY_{s}\sin(\beta_s h) & \cos(\beta_s h)
\end{bmatrix},
\end{align}
where $Z_s=1/Y_s=\sqrt{\mu_{s}/\epsilon_{s}}$ is the substrate impedance, and $\beta_s=\omega\sqrt{\mu_s\epsilon_s}$, where $\epsilon_{s}$ is the dielectric permittivity and $\mu_{s}$ is the permeability. Transformation of the $ABCD$ transfer matrix \textit{\textbf{M$_{ms}$}} into other network representations such as the scattering matrix \textit{\textbf{S$_{ms}$}} can be performed, through which we obtain $T=S_{21}$ as the metasurface transmission and $R=S_{11}$ as the metasurface reflection, written as \cite{pozar2011microwave}
\begin{align}
T&=\frac{2}{(A_{ms}+(B_{ms}/\eta)+(\eta C_{ms})+D_{ms})} \\
R&=\frac{(A_{ms}+(B_{ms}/\eta)-(\eta C_{ms})-D_{ms})}{A_{ms}+(B_{ms}/\eta)+(\eta C_{ms})+D_{ms}},
\end{align}
where $A_{ms}, B_{ms}, C_{ms}$, and $D_{ms}$ are elements of matrix \textit{\textbf{M$_{ms}$}} in Eq.~\eqref{eq:m_ms}.

\section{Calculation of the DC Admittance \label{apx:C}}

To derive the DC admittance as presented in Eq.~\eqref{eq:Yse_t}, we use transient analysis for the corresponding DC circuits. First, let us consider an $RL$ series metasurface, in which the embedded circuit is equivalent to a series configuration of an $RL$ circuit and $R_d$ that represents the resistive components of two diodes at their turn-on voltage \cite{asano2020simplified} (see Fig.~\ref{fig:2}(c)).  In this circuit, using the Kirchhoff voltage law (KVL), we find that initial potential $E_0$ is equivalent to the sum of all transient voltages within the resistors and inductor, i.e.,
\begin{equation} \label{eq:transient_RL}
    E_0=(R_L+R_d)i(t)+L\frac{di(t)}{dt}. 
\end{equation}
Hence, solving for circuit current $i(t)$ using the first-order differential equation and rearranging it, we obtain the DC admittance as \cite{asano2020simplified}
\begin{equation} \label{eq:Ydc_L}
    \frac{i(t)}{E_0}=\frac{1}{R_L+R_d}\left(1-e^{\left(-\frac{R_L+R_d}{L}\right)t}\right).
\end{equation}

Second, we consider an $RC$ parallel circuit, as seen in Fig.~\ref{fig:2}(d). In this circuit, we can use the Kirchhoff current law (KCL) to obtain the transient current as follows:
\begin{equation}
    i(t)=i_{RC}(t)+i_C(t),
    \end{equation}
Where $i_{RC}$ and $i_{C}$ represent the currents of $R$ and $C$, respectively. Using the first-order differential equation, we can derive $i_C$ and $i_{RC}$ as follows \cite{asano2020simplified}:
\begin{equation}
\begin{aligned}
i_C(t)=\frac{E_0}{R_d}e^{-\frac{R_C+R_d}{CR_CR_d}t}, \\
    i_{RC}(t)=\frac{E_0\left(1-e^{-\frac{R_C+R_d}{CR_CR_d}t}\right)}{R_d+R_C}.
\end{aligned}
\end{equation}
Hence, the DC admittance is
\begin{equation} \label{eq:Ydc_C}
    \frac{i(t)}{E_0}=\frac{i_{RC}(t)+i_C(t)}{E_0}.
\end{equation}

Next, we consider the series and parallel $RLC$ circuits seen in Fig.~\ref{fig:2}(e) and (f). In contrast to our previous study \cite{asano2020simplified}, in these circuits, the transient currents can be readily calculated by removing the $R_C$ from Fig.~\ref{fig:2}(d) and $R_L$ from Fig.~\ref{fig:2}(e), assuming that both $R_C$ and $R_L$ have small contributions to the overall transient response (see Appendix~\ref{apx:D}). In the series $RLC$, we use the KVL to obtain the following relationship:
\begin{equation}
    E_0=(R_d+R_L)i(t)+L\frac{di(t)}{dt}+\frac{1}{C}\int i(t)dt.
\end{equation}
Note that here, we do not have $R_C$ connected in parallel to $C$. In differential equation form, using $s$ to represent the roots, the above formula can be written as
\begin{equation} \label{eq:diff_eq_series}
    \left(s^2+\frac{R_d+R_L}{L}s+\frac{1}{LC}\right)i=0.
\end{equation}
The solution of this second-order, linear, homogenous differential equation is then \begin{equation}
    i(t)=K_1e^{s_1 t}+K_2e^{s_2 t},
\end{equation}
where $s_1$ and $s_2$ are the roots of the differential equation in Eq.~\eqref{eq:diff_eq_series}, and $K_1$ and $K_2$ are two constants that are to be selected to satisfy the initial conditions. First, at $t=0$, the initial condition is $e^{s_1 t}=e^{s_2 t}=1$; hence, $0=K_1+K_2$. Second, as the current in the inductor cannot change instantaneously, the initial condition of the voltage at $t=0^+$ across the inductor is the same as $E_0$, i.e., $L\frac{di}{dt}=E_0$. Hence, we obtain $L(s_1K_1+s_2K_2)=E_0$. 
Substituting these two equations, we obtain
$K_2=\frac{E_0}{L(s_2-s_1)}$ and $K_1=-\frac{E_0}{L(s_2-s_1)}$, and therefore, the DC admittance is derived from 
\begin{equation} \label{eq:Ydc_series}
    \frac{i(t)}{E_0}=\frac{1}{L(s_2-s_1)}\big(e^{s_1 t}-e^{s_2 t}\big).
\end{equation}

For the parallel RLC circuit, a similar derivation as for the series RLC circuit can be conducted. Here, application of the KCL to the parallel $RLC$ circuit yields
\begin{equation}
    \frac{E_0-v_p}{R_d}=\frac{v_p(t)}{R_C}+C\frac{dV_p(t)}{dt}+\frac{1}{L}\int v_p(t)dt,
\end{equation}
where $v_p$ is the voltage applied to the entire circuit except for $R_d$ (see Fig.~\ref{fig:2}(f)). In differential equation form, using $p$ to represent the roots, the above formula can be written as
\begin{equation}
    \left(p^2+\frac{R_d+R_C}{CR_CR_d}p+\frac{1}{LC}\right)v_p=0.
\end{equation}
The solution of this second-order, linear, homogenous differential equation is %
\begin{equation}
    v_p(t)=\kappa_1e^{p_1 t}+\kappa_2e^{p_2 t},
\end{equation}
where $p_1$ and $p_2$ are the roots of the above differential equation, and $\kappa_1$ and $\kappa_2$ are two constants that are to be selected to satisfy the initial conditions. First, at $t=0$, the initial condition is $e^{p_1 t}=e^{p_2 t}=1$; hence,
$0=\kappa_1+\kappa_2$.
Second, as the voltage in the capacitor cannot change instantaneously, the initial condition of the current at $t=0^+$ across the capacitor is same as initial current $i(t)=\frac{E_0-v_p}{R_d}$, i.e., $C\frac{dv_p}{dt}=\frac{E_0}{R_d}$. Hence, we obtain $R_dC(p_1\kappa_1+p_2\kappa_2)=E_0$.
Substituting these two equations, we obtain
$\kappa_1=\frac{E_0}{R_dC(p_2-p_1)}$ and $    \kappa_2=-\frac{E_0}{R_dC(p_2-p_1)}$, and hence, the DC admittance is
\begin{equation}
\begin{aligned} \label{eq:Ydc_parallel}
    \frac{i(t)}{E_0}=\frac{1}{R_d}-\frac{1}{R_d^2C(p_2-p_1)}\big(e^{p_1 t}-e^{p_2 t}\big).
\end{aligned}
\end{equation}

To obtain the analytically calculated complex admittance in Eq.~\eqref{eq:Yse_tf}, we use the following method. First, we derive the equivalent circuit of the metallic slit based on the procedure described in \cite{olk2019accurate, fathnan2020bandwidth}. This extraction requires full-wave simulation of the slit structure without the embedded circuit.  From the extraction, we obtain the equivalent $LC$ circuit for the metallic slit, which, in this case, consists of $C_m$=0.12\,pF and $L_m$=3.15\,nH (see Eq.~\eqref{eq:Yse_f}). Note that here, the metallic slit geometries are identical to those used in Fig.~\ref{fig:2}(a). Through Eq.~\eqref{eq:Yse_f}, we can obtain the frequency-dependent admittance of the slit structure utilizing $C_m$ and $L_m$. We then take the admittance only at the considered frequency (4.2\,GHz), which is to be added to the time-varying admittance in Eq.~\eqref{eq:Yse_t}. Here, after obtaining the DC admittance (either from Eqs.~\eqref{eq:Ydc_L},~\eqref{eq:Ydc_C},~\eqref{eq:Ydc_series},~or~\eqref{eq:Ydc_parallel}), the adjustment factors $\dot{a}$ and $\dot{b}$ are chosen to fit the corresponding numerical simulation results. This fitting is conducted only \emph{once} for a particular transient circuit type, and therefore is valid to be used in predicting the admittance of other circuit components within the same circuit type. For example, in the $RL$ series circuit, we adjust $\dot{a}$ and $\dot{b}$ to fit the simulation result of the metasurface with inductor $L$=0.01\,mH. Once obtained, further alteration of the adjustment factors is not required, even when the inductor changes to $L=$0.1\,mH and $L=$1\,mH. The elements of $\dot{a}$ and $\dot{b}$ for the four different circuits considered in this simulation are presented in Table~\ref{tab:adjustment_factors}. 

\begin{table}[h]
\caption{Adjustment factors for $\dot{a}$ and $\dot{b}$ of Eqs.~\eqref{eq:Yse_f} and \eqref{eq:Yse_t}}
\begin{ruledtabular}
\begin{tabular}{lllll}
              Circuit Type& $a'$   & $a''$   & $b'$   & $b''$    \\ \hline
\multicolumn{1}{l}{$RL$ Series and $RLC$ Series} & 0.1 & 2  & 2.7 & 2.7     \\ \hline
\multicolumn{1}{l}{$RC$ Parallel and $RLC$ Parallel} & 0.3 & 2.15 & 2.7 & 1.8  \\ 
\end{tabular}
\end{ruledtabular} \label{tab:adjustment_factors}
\end{table}

\section{DC Circuit Simulation Validating the Analytical Calculation \label{apx:D}}

\begin{figure}[t]
    \centering
    \includegraphics[width=0.9\linewidth]{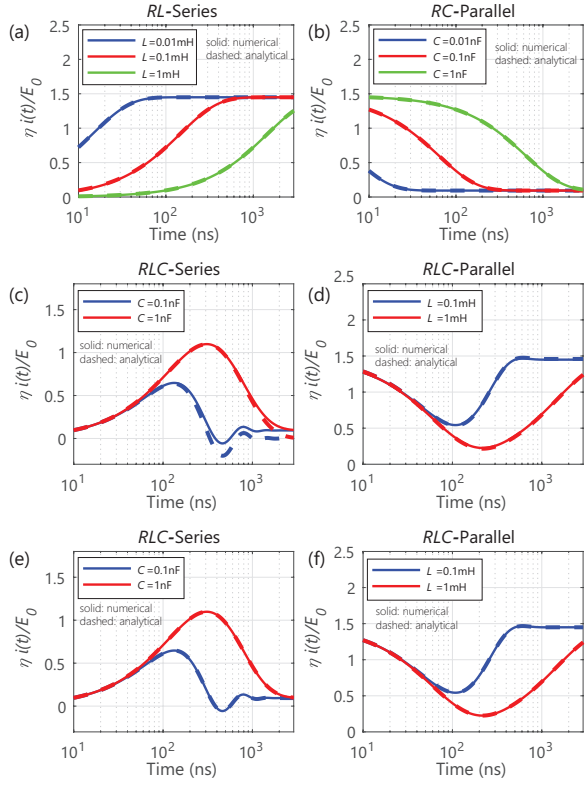}
    \caption{Comparison between the circuit simulation and analytically calculated DC admittance as explained in Appendix~\ref{apx:C} for (a) $RL$ series, (b) $RC$ parallel, (c) $RLC$ series, and (d) $RLC$ parallel. Comparison with the analytical expression of DC admittances from \cite{asano2020simplified} for (e) $RLC$ series and (f) $RLC$ parallel. In the above simulations and analytical calculations, $R_L=10\,\Omega$ and $R_C=10\,$k$\Omega$.}
    \label{fig:apx2}
\end{figure}

After obtaining an analytical expression of the DC admittances for the four different circuit types, i.e., Eqs.~\eqref{eq:Ydc_L},~\eqref{eq:Ydc_C},~\eqref{eq:Ydc_series},~and~\eqref{eq:Ydc_parallel}, we use the circuit simulator inside ANSYS Electronics Desktop 2020 R2 to compare the results. As can be seen from Fig.~\ref{fig:apx2}(a)-(d), both the analytically calculated DC admittances and numerical results from the circuit simulator agree with each other. In Fig.~\ref{fig:apx2}(c)-(d), a discrepancy can be found starting around 500\,ns due to the simplification used in the $RLC$ series and parallel circuits.
As discussed in Appendix~\ref{apx:C}, $R_L$ and $R_C$ are removed from the $RLC$ series case and the $RLC$ parallel case, respectively, to propose the simplified yet realistic calculation method of the DC admittances, assuming that both resistances only slightly affect the overall admittance. This assumption remains valid for $R_L$ approaching zero, indicating a closed connection, and $R_C$ approaching infinity, indicating an open connection. With $R_L=10\,\Omega$ and $R_C=10\,$k$\Omega$ used in this study, both the $RLC$ series and parallel admittances remain largely consistent. 
In \cite{asano2020simplified}, the DC admittance was derived considering both $R_L$ and $R_C$ within the circuit, which gives results more consistent with those of the circuit simulator, as can be seen from Fig.~\ref{fig:apx2}(e) and (f). In this calculation, we obtain $v_p$ as the voltage across the circuit of the $RLC$ parallel case, except for the diodes, as follows \cite{asano2020simplified}:

\footnotesize
\begin{multline}  \label{eq:v_p}
   v_p\ (t)=R_p E_0\bigg[R_l\bigg( {1+\frac{\beta e^{\alpha t}-\alpha e^{\beta t}}{\alpha-\beta}}\bigg)+L\frac{\alpha\beta(e^{\alpha t}-e^{\beta t})}{\alpha-\beta}\ \bigg],
\end{multline}
\normalsize
where $R_p=\frac{R_c}{R_l R_d+R_l R_c+R_d R_c}$, 
and $\alpha$ and $\beta$ are roots of the differential equations derived in \cite{asano2020simplified}, which are rewritten here as
\footnotesize
\begin{multline}
    \alpha=-\frac{L(R_d+R_c)+CR_l R_d R_c}{2LCR_d R_c}\\+\sqrt{\bigg({\frac{L(R_d+R_c )+CR_l R_d R_c}{2LCR_d R_c}\bigg)}^2-\frac{1}{LCR_d R_p}},
\end{multline}
\begin{multline}
    \beta=-\frac{L(R_d+R_c)+CR_l R_d R_c}{2LCR_d R_c}\\-\sqrt{\bigg({\frac{L(R_d+R_c )+CR_l R_d R_c}{2LCR_d R_c}\bigg)}^2-\frac{1}{LCR_d R_p}}.
\end{multline}
\normalsize

\begin{SCfigure*}
    \centering
    \includegraphics[width=1.57\linewidth]{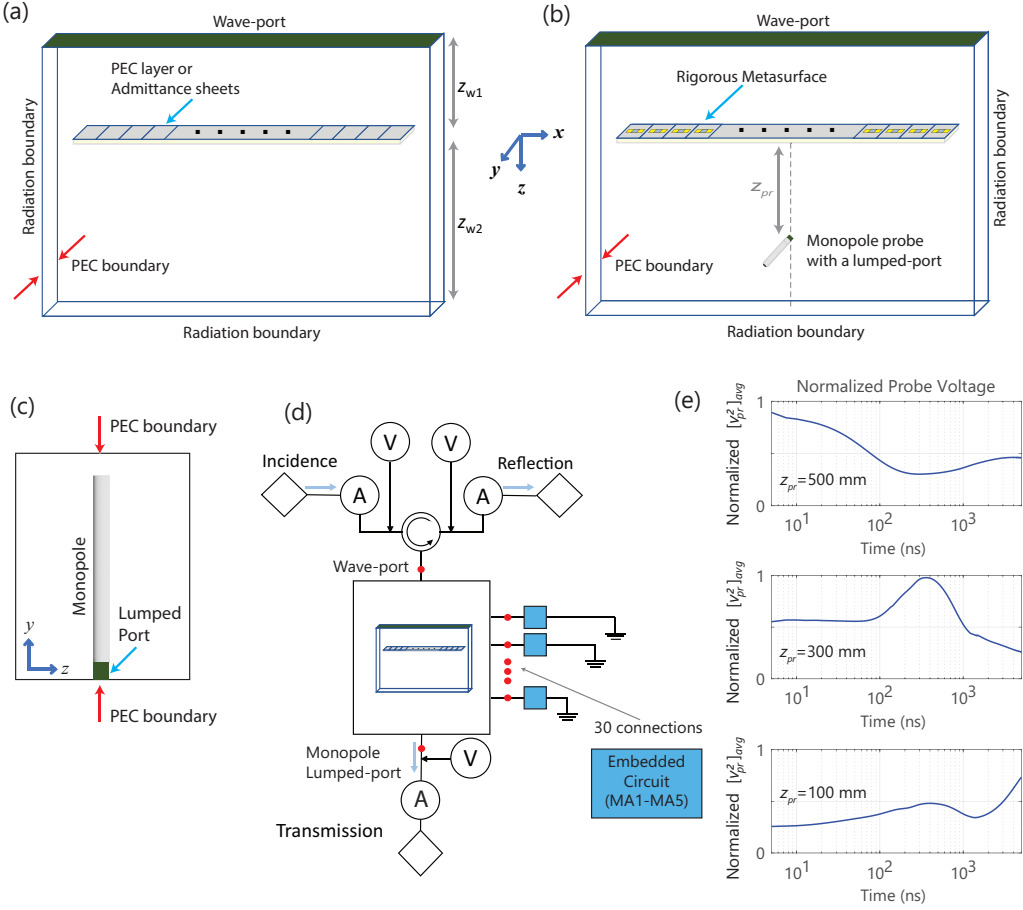}
    \caption{Simulation method for the FZPs based on (a) ideal PEC layers or equivalent admittance sheets and (b) rigorous waveform-selective metasurface. In these simulations, only 1D horizontal profiles of the FZPs are considered by applying PEC boundaries to both positive and negative $y$-axis directions (red arrows in (a) and (b)). For the FZP based on the admittance sheets, the metasurface is approximated by impedance boundaries on top of a dielectric slab. For the full FZP based on the rigorous waveform-selective metasurface, a monopole probe is used to capture the time-varying voltage. (c) The monopole placement within the metasurface scattering area. (d) Schematic of the circuit simulation of the full FZP. (e) Normalized squared and averaged time-varying voltages ($[v_{pr}^2]_{avg}$) for different probe locations ($z_{pr}$).}
    \label{fig:apx3}
\end{SCfigure*}

In the $RLC$ series case, considering that $R_C$ is connected in parallel to the capacitor $C$ (see Fig.~\ref{fig:2}(e)), we obtain $v_s$ as the voltage within the circuit, except for the diodes, as follows \cite{asano2020simplified}:
\footnotesize
\begin{multline} \label{eq:v_s}
    v_s(t)=\frac{E_0}{R_d+R_l+R_c} \bigg[(R_c+R_l){1-e^{ht}(\cos kt-\frac{h}{k} \sin kt)} \\ 
    +\frac{h^2+k^2}{k} e^{ht} {(C R_c R_l + L(1+C R_c h)) \sin kt + L C R_c k\cos kt}\bigg],
\end{multline}
\normalsize
in which the real and imaginary roots $h$ and $k$ are
\footnotesize
\begin{equation}
    h=-\frac{L+CR_c(R_d+R_l)}{2 L C R_c},
\end{equation}
\begin{equation}
    k=\sqrt{\frac{R_d+R_l+R_c}{LCR_c}-{\frac{L+CR_c(R_d+R_l)}{2LCR_c}}^2}.
\end{equation}
\normalsize

More information is provided in Ref.~\cite{asano2020simplified} for $v_p$ and $v_s$ which have two formulations depending on the combination of $RLC$ parameters. For simplicity here we only show one version of $v_p$ and $v_s$. We confirmed that both Eqs.~\eqref{eq:v_p} and \eqref{eq:v_s} can accurately calculate the admittance within the considered $RLC$ parameters as seen in Fig.~\ref{fig:apx2}(e) and (f).

\section{FZP Simulations \label{apx:E}}
The simulation environment for the FZPs based on ideal PEC layers or equivalent admittance sheets is depicted in Fig.~\ref{fig:apx3}(a), while that for the FZP based on rigorous waveform-selective metasurface is shown in Fig.~\ref{fig:apx3}(b). 
The simulations are performed only at the operating frequency of 4.2\,GHz. In these simulations, we only consider 1D horizontal profiles of the FZPs along the \textit{x}-axis by applying PEC boundaries to the \textit{y}-axis directions. Waveguide port excitation is assigned with one of the \textit{z}-axis directions with a minimum distance of $z_{w1}=200$\,mm (${\sim}3\lambda$) from the metasurface. All the other boundaries are set to be radiation boundaries. The simulation area in front of the metasurface (positive $z-$axis) is set to a distance $z_{w2}=800$\,mm (${\sim}11\lambda$). 

In the admittance sheet model, impedance boundary conditions are used to approximate the time-varying admittance only for three discretized instants. This simulation is similar to the single meta-atom simulation as explained in Subsec.~\ref{sec:full-wave MA admittance}. However, here the metasurface is modeled to have inhomogeneous scattering profiles over the horizontal \textit{x}-axis. 
Using this full-wave simulation of the admittance sheet model, we obtain scattered field profiles where a focusing effect can be observed (see Fig.~\ref{fig:7}(a) and (b)). 

In the full FZP based on the rigorous waveform-selective metasurface, the co-simulation method explained in Appendix~\ref{apx:A} is used. The meta-atoms here are realistic metallic (PEC) slits  with lumped ports. Note that in the full-wave simulation there is no variation in either the lumped ports or the slit dimensions. The variation to realize the inhomogeneous temporal profile is made inside the circuit simulation by implementing the corresponding circuit configuration, i.e., MA1 to MA5, where the detailed circuit parameters are presented in Table~\ref{tab:MA_parameters}. Although the full FZP consists of 42 cells in total, the center 12 cells have no temporal variation and thus are left blank. To observe the time-varying scattered field, a monopole probe is used with a variable $z$-axis coordinate at the center of the metasurface since the co-simulation method adopted cannot produce the transient field profile. As illustrated in Fig.~\ref{fig:apx3}(c), the monopole is a metallic (copper) cylinder with a length of 16\,mm and a diameter of 1\,mm. At one of its corner, a lumped port (1\,mm $\times$ 1\,mm) is connected. The circuit simulation of the full FZP is depicted in Fig.~\ref{fig:apx3}(d). The electromagnetic simulation result is imported to the circuit simulator of ANSYS Electronics Desktop (2020 R2). After the necessary components are connected via the lumped ports, we run a transient simulation over a 5-$\mu$s time span. Note that the input power here is set to 30\,dBm to ensure that the incident fields generate a sufficiently large potential across diodes so that the diodes are turned on. The voltage received by the monopole probe ($v_{pr}$) is calculated for three distances corresponding to three simulations with $z_{pr}=$500, 300, and 100\,mm. The voltages are squared and averaged ($[v_{pr}^2]_{avg}$) over 20 times wave period ($20\times 0.24$\,ns) in Fig.~\ref{fig:apx3}(e) and are normalized to be compared with the results in Fig.~\ref{fig:7}(c).


\nocite{*}
\bibliography{References}

\begin{thebibliography}{61}%
\makeatletter
\providecommand \@ifxundefined [1]{%
 \@ifx{#1\undefined}
}%
\providecommand \@ifnum [1]{%
 \ifnum #1\expandafter \@firstoftwo
 \else \expandafter \@secondoftwo
 \fi
}%
\providecommand \@ifx [1]{%
 \ifx #1\expandafter \@firstoftwo
 \else \expandafter \@secondoftwo
 \fi
}%
\providecommand \natexlab [1]{#1}%
\providecommand \enquote  [1]{``#1''}%
\providecommand \bibnamefont  [1]{#1}%
\providecommand \bibfnamefont [1]{#1}%
\providecommand \citenamefont [1]{#1}%
\providecommand \href@noop [0]{\@secondoftwo}%
\providecommand \href [0]{\begingroup \@sanitize@url \@href}%
\providecommand \@href[1]{\@@startlink{#1}\@@href}%
\providecommand \@@href[1]{\endgroup#1\@@endlink}%
\providecommand \@sanitize@url [0]{\catcode `\\12\catcode `\$12\catcode
  `\&12\catcode `\#12\catcode `\^12\catcode `\_12\catcode `\%12\relax}%
\providecommand \@@startlink[1]{}%
\providecommand \@@endlink[0]{}%
\providecommand \url  [0]{\begingroup\@sanitize@url \@url }%
\providecommand \@url [1]{\endgroup\@href {#1}{\urlprefix }}%
\providecommand \urlprefix  [0]{URL }%
\providecommand \Eprint [0]{\href }%
\providecommand \doibase [0]{https://doi.org/}%
\providecommand \selectlanguage [0]{\@gobble}%
\providecommand \bibinfo  [0]{\@secondoftwo}%
\providecommand \bibfield  [0]{\@secondoftwo}%
\providecommand \translation [1]{[#1]}%
\providecommand \BibitemOpen [0]{}%
\providecommand \bibitemStop [0]{}%
\providecommand \bibitemNoStop [0]{.\EOS\space}%
\providecommand \EOS [0]{\spacefactor3000\relax}%
\providecommand \BibitemShut  [1]{\csname bibitem#1\endcsname}%
\let\auto@bib@innerbib\@empty
\bibitem [{\citenamefont {Alu}\ \emph {et~al.}(2007)\citenamefont {Alu},
  \citenamefont {Silveirinha}, \citenamefont {Salandrino},\ and\ \citenamefont
  {Engheta}}]{alu2007epsilon}%
  \BibitemOpen
  \bibfield  {author} {\bibinfo {author} {\bibfnamefont {A.}~\bibnamefont
  {Alu}}, \bibinfo {author} {\bibfnamefont {M.~G.}\ \bibnamefont
  {Silveirinha}}, \bibinfo {author} {\bibfnamefont {A.}~\bibnamefont
  {Salandrino}},\ and\ \bibinfo {author} {\bibfnamefont {N.}~\bibnamefont
  {Engheta}},\ }\bibfield  {title} {\bibinfo {title} {Epsilon-near-zero
  metamaterials and electromagnetic sources: Tailoring the radiation phase
  pattern},\ }\href@noop {} {\bibfield  {journal} {\bibinfo  {journal} {Phys.
  Rev. B}\ }\textbf {\bibinfo {volume} {75}},\ \bibinfo {pages} {155410}
  (\bibinfo {year} {2007})}\BibitemShut {NoStop}%
\bibitem [{\citenamefont {Pfeiffer}\ and\ \citenamefont
  {Grbic}(2013)}]{pfeiffer2013metamaterial}%
  \BibitemOpen
  \bibfield  {author} {\bibinfo {author} {\bibfnamefont {C.}~\bibnamefont
  {Pfeiffer}}\ and\ \bibinfo {author} {\bibfnamefont {A.}~\bibnamefont
  {Grbic}},\ }\bibfield  {title} {\bibinfo {title} {Metamaterial huygens’
  surfaces: tailoring wave fronts with reflectionless sheets},\ }\href@noop {}
  {\bibfield  {journal} {\bibinfo  {journal} {Phys. Rev. Lett.}\ }\textbf
  {\bibinfo {volume} {110}},\ \bibinfo {pages} {197401} (\bibinfo {year}
  {2013})}\BibitemShut {NoStop}%
\bibitem [{\citenamefont {Ding}\ \emph {et~al.}(2015)\citenamefont {Ding},
  \citenamefont {Monticone}, \citenamefont {Zhang}, \citenamefont {Zhang},
  \citenamefont {Gao}, \citenamefont {Burokur}, \citenamefont {de~Lustrac},
  \citenamefont {Wu}, \citenamefont {Qiu},\ and\ \citenamefont
  {Alu}}]{ding2015ultrathin}%
  \BibitemOpen
  \bibfield  {author} {\bibinfo {author} {\bibfnamefont {X.}~\bibnamefont
  {Ding}}, \bibinfo {author} {\bibfnamefont {F.}~\bibnamefont {Monticone}},
  \bibinfo {author} {\bibfnamefont {K.}~\bibnamefont {Zhang}}, \bibinfo
  {author} {\bibfnamefont {L.}~\bibnamefont {Zhang}}, \bibinfo {author}
  {\bibfnamefont {D.}~\bibnamefont {Gao}}, \bibinfo {author} {\bibfnamefont
  {S.~N.}\ \bibnamefont {Burokur}}, \bibinfo {author} {\bibfnamefont
  {A.}~\bibnamefont {de~Lustrac}}, \bibinfo {author} {\bibfnamefont
  {Q.}~\bibnamefont {Wu}}, \bibinfo {author} {\bibfnamefont {C.-W.}\
  \bibnamefont {Qiu}},\ and\ \bibinfo {author} {\bibfnamefont {A.}~\bibnamefont
  {Alu}},\ }\bibfield  {title} {\bibinfo {title} {Ultrathin pancharatnam--berry
  metasurface with maximal cross-polarization efficiency},\ }\href@noop {}
  {\bibfield  {journal} {\bibinfo  {journal} {Adv. Mater.}\ }\textbf {\bibinfo
  {volume} {27}},\ \bibinfo {pages} {1195} (\bibinfo {year}
  {2015})}\BibitemShut {NoStop}%
\bibitem [{\citenamefont {Yuan}\ \emph {et~al.}(2020)\citenamefont {Yuan},
  \citenamefont {Sun}, \citenamefont {Chen}, \citenamefont {Zhang},
  \citenamefont {Ding}, \citenamefont {Ratni}, \citenamefont {Wu},
  \citenamefont {Burokur},\ and\ \citenamefont {Qiu}}]{yuan2020fully}%
  \BibitemOpen
  \bibfield  {author} {\bibinfo {author} {\bibfnamefont {Y.}~\bibnamefont
  {Yuan}}, \bibinfo {author} {\bibfnamefont {S.}~\bibnamefont {Sun}}, \bibinfo
  {author} {\bibfnamefont {Y.}~\bibnamefont {Chen}}, \bibinfo {author}
  {\bibfnamefont {K.}~\bibnamefont {Zhang}}, \bibinfo {author} {\bibfnamefont
  {X.}~\bibnamefont {Ding}}, \bibinfo {author} {\bibfnamefont {B.}~\bibnamefont
  {Ratni}}, \bibinfo {author} {\bibfnamefont {Q.}~\bibnamefont {Wu}}, \bibinfo
  {author} {\bibfnamefont {S.~N.}\ \bibnamefont {Burokur}},\ and\ \bibinfo
  {author} {\bibfnamefont {C.-W.}\ \bibnamefont {Qiu}},\ }\bibfield  {title}
  {\bibinfo {title} {A fully phase-modulated metasurface as an
  energy-controllable circular polarization router},\ }\href@noop {} {\bibfield
   {journal} {\bibinfo  {journal} {Advanced Science}\ }\textbf {\bibinfo
  {volume} {7}},\ \bibinfo {pages} {2001437} (\bibinfo {year}
  {2020})}\BibitemShut {NoStop}%
\bibitem [{\citenamefont {Zhang}\ \emph {et~al.}(2021)\citenamefont {Zhang},
  \citenamefont {Wang}, \citenamefont {Burokur},\ and\ \citenamefont
  {Wu}}]{zhang2021generating}%
  \BibitemOpen
  \bibfield  {author} {\bibinfo {author} {\bibfnamefont {K.}~\bibnamefont
  {Zhang}}, \bibinfo {author} {\bibfnamefont {Y.}~\bibnamefont {Wang}},
  \bibinfo {author} {\bibfnamefont {S.~N.}\ \bibnamefont {Burokur}},\ and\
  \bibinfo {author} {\bibfnamefont {Q.}~\bibnamefont {Wu}},\ }\bibfield
  {title} {\bibinfo {title} {Generating dual-polarized vortex beam by detour
  phase: From phase gradient metasurfaces to metagratings},\ }\href@noop {}
  {\bibfield  {journal} {\bibinfo  {journal} {IEEE Transactions on Microwave
  Theory and Techniques}\ }\textbf {\bibinfo {volume} {70}},\ \bibinfo {pages}
  {200} (\bibinfo {year} {2021})}\BibitemShut {NoStop}%
\bibitem [{\citenamefont {Ramaccia}\ \emph {et~al.}(2019)\citenamefont
  {Ramaccia}, \citenamefont {Sounas}, \citenamefont {Alu}, \citenamefont
  {Toscano},\ and\ \citenamefont {Bilotti}}]{ramaccia2019phase}%
  \BibitemOpen
  \bibfield  {author} {\bibinfo {author} {\bibfnamefont {D.}~\bibnamefont
  {Ramaccia}}, \bibinfo {author} {\bibfnamefont {D.~L.}\ \bibnamefont
  {Sounas}}, \bibinfo {author} {\bibfnamefont {A.}~\bibnamefont {Alu}},
  \bibinfo {author} {\bibfnamefont {A.}~\bibnamefont {Toscano}},\ and\ \bibinfo
  {author} {\bibfnamefont {F.}~\bibnamefont {Bilotti}},\ }\bibfield  {title}
  {\bibinfo {title} {Phase-induced frequency conversion and doppler effect with
  time-modulated metasurfaces},\ }\href@noop {} {\bibfield  {journal} {\bibinfo
   {journal} {IEEE Trans. Antennas Propag.}\ }\textbf {\bibinfo {volume}
  {68}},\ \bibinfo {pages} {1607} (\bibinfo {year} {2019})}\BibitemShut
  {NoStop}%
\bibitem [{\citenamefont {Hadad}\ \emph {et~al.}(2015)\citenamefont {Hadad},
  \citenamefont {Sounas},\ and\ \citenamefont {Alu}}]{hadad2015space}%
  \BibitemOpen
  \bibfield  {author} {\bibinfo {author} {\bibfnamefont {Y.}~\bibnamefont
  {Hadad}}, \bibinfo {author} {\bibfnamefont {D.~L.}\ \bibnamefont {Sounas}},\
  and\ \bibinfo {author} {\bibfnamefont {A.}~\bibnamefont {Alu}},\ }\bibfield
  {title} {\bibinfo {title} {Space-time gradient metasurfaces},\ }\href@noop {}
  {\bibfield  {journal} {\bibinfo  {journal} {Phys. Rev. B}\ }\textbf {\bibinfo
  {volume} {92}},\ \bibinfo {pages} {100304} (\bibinfo {year}
  {2015})}\BibitemShut {NoStop}%
\bibitem [{\citenamefont {Taravati}\ and\ \citenamefont
  {Eleftheriades}(2019)}]{taravati2019generalized}%
  \BibitemOpen
  \bibfield  {author} {\bibinfo {author} {\bibfnamefont {S.}~\bibnamefont
  {Taravati}}\ and\ \bibinfo {author} {\bibfnamefont {G.~V.}\ \bibnamefont
  {Eleftheriades}},\ }\bibfield  {title} {\bibinfo {title} {Generalized
  space-time-periodic diffraction gratings: Theory and applications},\
  }\href@noop {} {\bibfield  {journal} {\bibinfo  {journal} {Phys. Rev. App.}\
  }\textbf {\bibinfo {volume} {12}},\ \bibinfo {pages} {024026} (\bibinfo
  {year} {2019})}\BibitemShut {NoStop}%
\bibitem [{\citenamefont {Zhang}\ \emph {et~al.}(2019)\citenamefont {Zhang},
  \citenamefont {Chen}, \citenamefont {Shao}, \citenamefont {Dai},
  \citenamefont {Cheng}, \citenamefont {Castaldi}, \citenamefont {Galdi},\ and\
  \citenamefont {Cui}}]{zhang2019breaking}%
  \BibitemOpen
  \bibfield  {author} {\bibinfo {author} {\bibfnamefont {L.}~\bibnamefont
  {Zhang}}, \bibinfo {author} {\bibfnamefont {X.~Q.}\ \bibnamefont {Chen}},
  \bibinfo {author} {\bibfnamefont {R.~W.}\ \bibnamefont {Shao}}, \bibinfo
  {author} {\bibfnamefont {J.~Y.}\ \bibnamefont {Dai}}, \bibinfo {author}
  {\bibfnamefont {Q.}~\bibnamefont {Cheng}}, \bibinfo {author} {\bibfnamefont
  {G.}~\bibnamefont {Castaldi}}, \bibinfo {author} {\bibfnamefont
  {V.}~\bibnamefont {Galdi}},\ and\ \bibinfo {author} {\bibfnamefont {T.~J.}\
  \bibnamefont {Cui}},\ }\bibfield  {title} {\bibinfo {title} {Breaking
  reciprocity with space-time-coding digital metasurfaces},\ }\href@noop {}
  {\bibfield  {journal} {\bibinfo  {journal} {Adv. Mater.}\ }\textbf {\bibinfo
  {volume} {31}},\ \bibinfo {pages} {1904069} (\bibinfo {year}
  {2019})}\BibitemShut {NoStop}%
\bibitem [{\citenamefont {F.~Imani}\ and\ \citenamefont
  {Smith}(2020)}]{lImani:2020}%
  \BibitemOpen
  \bibfield  {author} {\bibinfo {author} {\bibfnamefont {M.}~\bibnamefont
  {F.~Imani}}\ and\ \bibinfo {author} {\bibfnamefont {D.~R.}\ \bibnamefont
  {Smith}},\ }\bibfield  {title} {\bibinfo {title} {Temporal microwave ghost
  imaging using a reconfigurable disordered cavity},\ }\href@noop {} {\bibfield
   {journal} {\bibinfo  {journal} {Appl. Phy. Lett.}\ }\textbf {\bibinfo
  {volume} {116}},\ \bibinfo {pages} {054102} (\bibinfo {year}
  {2020})}\BibitemShut {NoStop}%
\bibitem [{\citenamefont {Zhao}\ \emph {et~al.}(2019)\citenamefont {Zhao},
  \citenamefont {Yang}, \citenamefont {Dai}, \citenamefont {Cheng},
  \citenamefont {Li}, \citenamefont {Qi}, \citenamefont {Ke}, \citenamefont
  {Bai}, \citenamefont {Liu}, \citenamefont {Jin} \emph
  {et~al.}}]{zhao2019programmable}%
  \BibitemOpen
  \bibfield  {author} {\bibinfo {author} {\bibfnamefont {J.}~\bibnamefont
  {Zhao}}, \bibinfo {author} {\bibfnamefont {X.}~\bibnamefont {Yang}}, \bibinfo
  {author} {\bibfnamefont {J.~Y.}\ \bibnamefont {Dai}}, \bibinfo {author}
  {\bibfnamefont {Q.}~\bibnamefont {Cheng}}, \bibinfo {author} {\bibfnamefont
  {X.}~\bibnamefont {Li}}, \bibinfo {author} {\bibfnamefont {N.~H.}\
  \bibnamefont {Qi}}, \bibinfo {author} {\bibfnamefont {J.~C.}\ \bibnamefont
  {Ke}}, \bibinfo {author} {\bibfnamefont {G.~D.}\ \bibnamefont {Bai}},
  \bibinfo {author} {\bibfnamefont {S.}~\bibnamefont {Liu}}, \bibinfo {author}
  {\bibfnamefont {S.}~\bibnamefont {Jin}}, \emph {et~al.},\ }\bibfield  {title}
  {\bibinfo {title} {Programmable time-domain digital-coding metasurface for
  non-linear harmonic manipulation and new wireless communication systems},\
  }\href@noop {} {\bibfield  {journal} {\bibinfo  {journal} {Natl. Sci. Rev.}\
  }\textbf {\bibinfo {volume} {6}},\ \bibinfo {pages} {231} (\bibinfo {year}
  {2019})}\BibitemShut {NoStop}%
\bibitem [{\citenamefont {Hall}\ \emph {et~al.}(2021)\citenamefont {Hall},
  \citenamefont {Yessenov}, \citenamefont {Ponomarenko},\ and\ \citenamefont
  {Abouraddy}}]{hall2021space}%
  \BibitemOpen
  \bibfield  {author} {\bibinfo {author} {\bibfnamefont {L.~A.}\ \bibnamefont
  {Hall}}, \bibinfo {author} {\bibfnamefont {M.}~\bibnamefont {Yessenov}},
  \bibinfo {author} {\bibfnamefont {S.~A.}\ \bibnamefont {Ponomarenko}},\ and\
  \bibinfo {author} {\bibfnamefont {A.~F.}\ \bibnamefont {Abouraddy}},\
  }\bibfield  {title} {\bibinfo {title} {The space--time talbot effect},\
  }\href@noop {} {\bibfield  {journal} {\bibinfo  {journal} {APL Photonics}\
  }\textbf {\bibinfo {volume} {6}},\ \bibinfo {pages} {056105} (\bibinfo {year}
  {2021})}\BibitemShut {NoStop}%
\bibitem [{\citenamefont {Liu}\ \emph {et~al.}(2018)\citenamefont {Liu},
  \citenamefont {Powell}, \citenamefont {Zarate},\ and\ \citenamefont
  {Shadrivov}}]{liu2018huygens}%
  \BibitemOpen
  \bibfield  {author} {\bibinfo {author} {\bibfnamefont {M.}~\bibnamefont
  {Liu}}, \bibinfo {author} {\bibfnamefont {D.~A.}\ \bibnamefont {Powell}},
  \bibinfo {author} {\bibfnamefont {Y.}~\bibnamefont {Zarate}},\ and\ \bibinfo
  {author} {\bibfnamefont {I.~V.}\ \bibnamefont {Shadrivov}},\ }\bibfield
  {title} {\bibinfo {title} {Huygens’ metadevices for parametric waves},\
  }\href@noop {} {\bibfield  {journal} {\bibinfo  {journal} {Phys. Rev. X}\
  }\textbf {\bibinfo {volume} {8}},\ \bibinfo {pages} {031077} (\bibinfo {year}
  {2018})}\BibitemShut {NoStop}%
\bibitem [{\citenamefont {Shaltout}\ \emph {et~al.}(2019)\citenamefont
  {Shaltout}, \citenamefont {Lagoudakis}, \citenamefont {van~de Groep},
  \citenamefont {Kim}, \citenamefont {Vu{\v{c}}kovi{\'c}}, \citenamefont
  {Shalaev},\ and\ \citenamefont {Brongersma}}]{shaltout2019spatiotemporal}%
  \BibitemOpen
  \bibfield  {author} {\bibinfo {author} {\bibfnamefont {A.~M.}\ \bibnamefont
  {Shaltout}}, \bibinfo {author} {\bibfnamefont {K.~G.}\ \bibnamefont
  {Lagoudakis}}, \bibinfo {author} {\bibfnamefont {J.}~\bibnamefont {van~de
  Groep}}, \bibinfo {author} {\bibfnamefont {S.~J.}\ \bibnamefont {Kim}},
  \bibinfo {author} {\bibfnamefont {J.}~\bibnamefont {Vu{\v{c}}kovi{\'c}}},
  \bibinfo {author} {\bibfnamefont {V.~M.}\ \bibnamefont {Shalaev}},\ and\
  \bibinfo {author} {\bibfnamefont {M.~L.}\ \bibnamefont {Brongersma}},\
  }\bibfield  {title} {\bibinfo {title} {Spatiotemporal light control with
  frequency-gradient metasurfaces},\ }\href@noop {} {\bibfield  {journal}
  {\bibinfo  {journal} {Science}\ }\textbf {\bibinfo {volume} {365}},\ \bibinfo
  {pages} {374} (\bibinfo {year} {2019})}\BibitemShut {NoStop}%
\bibitem [{\citenamefont {Salary}\ \emph {et~al.}(2019)\citenamefont {Salary},
  \citenamefont {Farazi},\ and\ \citenamefont
  {Mosallaei}}]{salary2019dynamically}%
  \BibitemOpen
  \bibfield  {author} {\bibinfo {author} {\bibfnamefont {M.~M.}\ \bibnamefont
  {Salary}}, \bibinfo {author} {\bibfnamefont {S.}~\bibnamefont {Farazi}},\
  and\ \bibinfo {author} {\bibfnamefont {H.}~\bibnamefont {Mosallaei}},\
  }\bibfield  {title} {\bibinfo {title} {A dynamically modulated all-dielectric
  metasurface doublet for directional harmonic generation and manipulation in
  transmission},\ }\href@noop {} {\bibfield  {journal} {\bibinfo  {journal}
  {Adv. Opt. Mater.}\ }\textbf {\bibinfo {volume} {7}},\ \bibinfo {pages}
  {1900843} (\bibinfo {year} {2019})}\BibitemShut {NoStop}%
\bibitem [{\citenamefont {Sedeh}\ \emph {et~al.}(2020)\citenamefont {Sedeh},
  \citenamefont {Salary},\ and\ \citenamefont {Mosallaei}}]{sedeh2020time}%
  \BibitemOpen
  \bibfield  {author} {\bibinfo {author} {\bibfnamefont {H.~B.}\ \bibnamefont
  {Sedeh}}, \bibinfo {author} {\bibfnamefont {M.~M.}\ \bibnamefont {Salary}},\
  and\ \bibinfo {author} {\bibfnamefont {H.}~\bibnamefont {Mosallaei}},\
  }\bibfield  {title} {\bibinfo {title} {Time-varying optical vortices enabled
  by time-modulated metasurfaces},\ }\href@noop {} {\bibfield  {journal}
  {\bibinfo  {journal} {Nanophotonics}\ }\textbf {\bibinfo {volume} {9}},\
  \bibinfo {pages} {2957} (\bibinfo {year} {2020})}\BibitemShut {NoStop}%
\bibitem [{\citenamefont {Zhang}\ \emph {et~al.}(2022)\citenamefont {Zhang},
  \citenamefont {Chen}, \citenamefont {Hu}, \citenamefont {Zhao}, \citenamefont
  {Zhao}, \citenamefont {Jiang},\ and\ \citenamefont
  {Feng}}]{zhang2022spatiotemporal}%
  \BibitemOpen
  \bibfield  {author} {\bibinfo {author} {\bibfnamefont {N.}~\bibnamefont
  {Zhang}}, \bibinfo {author} {\bibfnamefont {K.}~\bibnamefont {Chen}},
  \bibinfo {author} {\bibfnamefont {Q.}~\bibnamefont {Hu}}, \bibinfo {author}
  {\bibfnamefont {J.}~\bibnamefont {Zhao}}, \bibinfo {author} {\bibfnamefont
  {J.}~\bibnamefont {Zhao}}, \bibinfo {author} {\bibfnamefont {T.}~\bibnamefont
  {Jiang}},\ and\ \bibinfo {author} {\bibfnamefont {Y.}~\bibnamefont {Feng}},\
  }\bibfield  {title} {\bibinfo {title} {Spatiotemporal metasurface to control
  electromagnetic wave scattering},\ }\href@noop {} {\bibfield  {journal}
  {\bibinfo  {journal} {Phys. Rev. App.}\ }\textbf {\bibinfo {volume} {17}},\
  \bibinfo {pages} {054001} (\bibinfo {year} {2022})}\BibitemShut {NoStop}%
\bibitem [{\citenamefont {Hu}\ \emph {et~al.}(2022)\citenamefont {Hu},
  \citenamefont {Zhao}, \citenamefont {Chen}, \citenamefont {Qu}, \citenamefont
  {Yang}, \citenamefont {Zhao}, \citenamefont {Jiang},\ and\ \citenamefont
  {Feng}}]{hu2022intelligent}%
  \BibitemOpen
  \bibfield  {author} {\bibinfo {author} {\bibfnamefont {Q.}~\bibnamefont
  {Hu}}, \bibinfo {author} {\bibfnamefont {J.}~\bibnamefont {Zhao}}, \bibinfo
  {author} {\bibfnamefont {K.}~\bibnamefont {Chen}}, \bibinfo {author}
  {\bibfnamefont {K.}~\bibnamefont {Qu}}, \bibinfo {author} {\bibfnamefont
  {W.}~\bibnamefont {Yang}}, \bibinfo {author} {\bibfnamefont {J.}~\bibnamefont
  {Zhao}}, \bibinfo {author} {\bibfnamefont {T.}~\bibnamefont {Jiang}},\ and\
  \bibinfo {author} {\bibfnamefont {Y.}~\bibnamefont {Feng}},\ }\bibfield
  {title} {\bibinfo {title} {An intelligent programmable omni-metasurface},\
  }\href@noop {} {\bibfield  {journal} {\bibinfo  {journal} {Laser \& Photonics
  Reviews}\ }\textbf {\bibinfo {volume} {16}},\ \bibinfo {pages} {2100718}
  (\bibinfo {year} {2022})}\BibitemShut {NoStop}%
\bibitem [{\citenamefont {Caloz}\ and\ \citenamefont
  {Deck-Leger}(2019)}]{caloz2019spacetime1}%
  \BibitemOpen
  \bibfield  {author} {\bibinfo {author} {\bibfnamefont {C.}~\bibnamefont
  {Caloz}}\ and\ \bibinfo {author} {\bibfnamefont {Z.-L.}\ \bibnamefont
  {Deck-Leger}},\ }\bibfield  {title} {\bibinfo {title} {Spacetime
  metamaterials—part ii: theory and applications},\ }\href@noop {} {\bibfield
   {journal} {\bibinfo  {journal} {IEEE Trans. Antennas Propag.}\ }\textbf
  {\bibinfo {volume} {68}},\ \bibinfo {pages} {1583} (\bibinfo {year}
  {2019})}\BibitemShut {NoStop}%
\bibitem [{\citenamefont {Caloz}\ and\ \citenamefont
  {Deck-L{\'e}ger}(2019)}]{caloz2019spacetime2}%
  \BibitemOpen
  \bibfield  {author} {\bibinfo {author} {\bibfnamefont {C.}~\bibnamefont
  {Caloz}}\ and\ \bibinfo {author} {\bibfnamefont {Z.-L.}\ \bibnamefont
  {Deck-L{\'e}ger}},\ }\bibfield  {title} {\bibinfo {title} {Spacetime
  metamaterials—part i: general concepts},\ }\href@noop {} {\bibfield
  {journal} {\bibinfo  {journal} {IEEE Trans. Antennas Propag.}\ }\textbf
  {\bibinfo {volume} {68}},\ \bibinfo {pages} {1569} (\bibinfo {year}
  {2019})}\BibitemShut {NoStop}%
\bibitem [{\citenamefont {Taravati}\ and\ \citenamefont
  {Eleftheriades}(2022)}]{taravati2022microwave}%
  \BibitemOpen
  \bibfield  {author} {\bibinfo {author} {\bibfnamefont {S.}~\bibnamefont
  {Taravati}}\ and\ \bibinfo {author} {\bibfnamefont {G.~V.}\ \bibnamefont
  {Eleftheriades}},\ }\bibfield  {title} {\bibinfo {title} {Microwave
  space-time-modulated metasurfaces},\ }\href@noop {} {\bibfield  {journal}
  {\bibinfo  {journal} {ACS Photonics}\ }\textbf {\bibinfo {volume} {9}},\
  \bibinfo {pages} {305} (\bibinfo {year} {2022})}\BibitemShut {NoStop}%
\bibitem [{\citenamefont {Tang}\ \emph {et~al.}(2019)\citenamefont {Tang},
  \citenamefont {Dai}, \citenamefont {Chen}, \citenamefont {Li}, \citenamefont
  {Cheng}, \citenamefont {Jin}, \citenamefont {Wong},\ and\ \citenamefont
  {Cui}}]{tang2019programmable}%
  \BibitemOpen
  \bibfield  {author} {\bibinfo {author} {\bibfnamefont {W.}~\bibnamefont
  {Tang}}, \bibinfo {author} {\bibfnamefont {J.~Y.}\ \bibnamefont {Dai}},
  \bibinfo {author} {\bibfnamefont {M.}~\bibnamefont {Chen}}, \bibinfo {author}
  {\bibfnamefont {X.}~\bibnamefont {Li}}, \bibinfo {author} {\bibfnamefont
  {Q.}~\bibnamefont {Cheng}}, \bibinfo {author} {\bibfnamefont
  {S.}~\bibnamefont {Jin}}, \bibinfo {author} {\bibfnamefont {K.-K.}\
  \bibnamefont {Wong}},\ and\ \bibinfo {author} {\bibfnamefont {T.~J.}\
  \bibnamefont {Cui}},\ }\bibfield  {title} {\bibinfo {title} {Programmable
  metasurface-based rf chain-free 8psk wireless transmitter},\ }\href@noop {}
  {\bibfield  {journal} {\bibinfo  {journal} {Electron. Lett.}\ }\textbf
  {\bibinfo {volume} {55}},\ \bibinfo {pages} {417} (\bibinfo {year}
  {2019})}\BibitemShut {NoStop}%
\bibitem [{\citenamefont {Li}\ \emph {et~al.}(2017{\natexlab{a}})\citenamefont
  {Li}, \citenamefont {Jun~Cui}, \citenamefont {Ji}, \citenamefont {Liu},
  \citenamefont {Ding}, \citenamefont {Wan}, \citenamefont {Bo~Li},
  \citenamefont {Jiang}, \citenamefont {Qiu},\ and\ \citenamefont
  {Zhang}}]{li2017electromagnetic}%
  \BibitemOpen
  \bibfield  {author} {\bibinfo {author} {\bibfnamefont {L.}~\bibnamefont
  {Li}}, \bibinfo {author} {\bibfnamefont {T.}~\bibnamefont {Jun~Cui}},
  \bibinfo {author} {\bibfnamefont {W.}~\bibnamefont {Ji}}, \bibinfo {author}
  {\bibfnamefont {S.}~\bibnamefont {Liu}}, \bibinfo {author} {\bibfnamefont
  {J.}~\bibnamefont {Ding}}, \bibinfo {author} {\bibfnamefont {X.}~\bibnamefont
  {Wan}}, \bibinfo {author} {\bibfnamefont {Y.}~\bibnamefont {Bo~Li}}, \bibinfo
  {author} {\bibfnamefont {M.}~\bibnamefont {Jiang}}, \bibinfo {author}
  {\bibfnamefont {C.-W.}\ \bibnamefont {Qiu}},\ and\ \bibinfo {author}
  {\bibfnamefont {S.}~\bibnamefont {Zhang}},\ }\bibfield  {title} {\bibinfo
  {title} {Electromagnetic reprogrammable coding-metasurface holograms},\
  }\href@noop {} {\bibfield  {journal} {\bibinfo  {journal} {Nat. Commun.}\
  }\textbf {\bibinfo {volume} {8}},\ \bibinfo {pages} {1} (\bibinfo {year}
  {2017}{\natexlab{a}})}\BibitemShut {NoStop}%
\bibitem [{\citenamefont {Denz}\ \emph {et~al.}(2010)\citenamefont {Denz},
  \citenamefont {Flach}, \citenamefont {Kivshar} \emph
  {et~al.}}]{denz2010nonlinearities}%
  \BibitemOpen
  \bibfield  {author} {\bibinfo {author} {\bibfnamefont {C.}~\bibnamefont
  {Denz}}, \bibinfo {author} {\bibfnamefont {S.}~\bibnamefont {Flach}},
  \bibinfo {author} {\bibfnamefont {Y.~S.}\ \bibnamefont {Kivshar}}, \emph
  {et~al.},\ }\href@noop {} {\emph {\bibinfo {title} {Nonlinearities in
  periodic structures and metamaterials}}},\ Vol.\ \bibinfo {volume} {150}\
  (\bibinfo  {publisher} {Springer Berlin, Heidelberg},\ \bibinfo {year}
  {2010})\BibitemShut {NoStop}%
\bibitem [{\citenamefont {Wakatsuchi}\ \emph
  {et~al.}(2013{\natexlab{a}})\citenamefont {Wakatsuchi}, \citenamefont {Kim},
  \citenamefont {Rushton},\ and\ \citenamefont
  {Sievenpiper}}]{wakatsuchi2013circuit}%
  \BibitemOpen
  \bibfield  {author} {\bibinfo {author} {\bibfnamefont {H.}~\bibnamefont
  {Wakatsuchi}}, \bibinfo {author} {\bibfnamefont {S.}~\bibnamefont {Kim}},
  \bibinfo {author} {\bibfnamefont {J.~J.}\ \bibnamefont {Rushton}},\ and\
  \bibinfo {author} {\bibfnamefont {D.~F.}\ \bibnamefont {Sievenpiper}},\
  }\bibfield  {title} {\bibinfo {title} {Circuit-based nonlinear metasurface
  absorbers for high power surface currents},\ }\href@noop {} {\bibfield
  {journal} {\bibinfo  {journal} {App. Phys. Lett.}\ }\textbf {\bibinfo
  {volume} {102}},\ \bibinfo {pages} {214103} (\bibinfo {year}
  {2013}{\natexlab{a}})}\BibitemShut {NoStop}%
\bibitem [{\citenamefont {Kiani}\ \emph
  {et~al.}(2020{\natexlab{a}})\citenamefont {Kiani}, \citenamefont {Momeni},
  \citenamefont {Tayarani},\ and\ \citenamefont {Ding}}]{kiani2020spatial}%
  \BibitemOpen
  \bibfield  {author} {\bibinfo {author} {\bibfnamefont {M.}~\bibnamefont
  {Kiani}}, \bibinfo {author} {\bibfnamefont {A.}~\bibnamefont {Momeni}},
  \bibinfo {author} {\bibfnamefont {M.}~\bibnamefont {Tayarani}},\ and\
  \bibinfo {author} {\bibfnamefont {C.}~\bibnamefont {Ding}},\ }\bibfield
  {title} {\bibinfo {title} {Spatial wave control using a self-biased nonlinear
  metasurface at microwave frequencies},\ }\href@noop {} {\bibfield  {journal}
  {\bibinfo  {journal} {Opt. Express}\ }\textbf {\bibinfo {volume} {28}},\
  \bibinfo {pages} {35128} (\bibinfo {year} {2020}{\natexlab{a}})}\BibitemShut
  {NoStop}%
\bibitem [{\citenamefont {Kiani}\ \emph
  {et~al.}(2020{\natexlab{b}})\citenamefont {Kiani}, \citenamefont {Tayarani},
  \citenamefont {Momeni}, \citenamefont {Rajabalipanah},\ and\ \citenamefont
  {Abdolali}}]{kiani2020self}%
  \BibitemOpen
  \bibfield  {author} {\bibinfo {author} {\bibfnamefont {M.}~\bibnamefont
  {Kiani}}, \bibinfo {author} {\bibfnamefont {M.}~\bibnamefont {Tayarani}},
  \bibinfo {author} {\bibfnamefont {A.}~\bibnamefont {Momeni}}, \bibinfo
  {author} {\bibfnamefont {H.}~\bibnamefont {Rajabalipanah}},\ and\ \bibinfo
  {author} {\bibfnamefont {A.}~\bibnamefont {Abdolali}},\ }\bibfield  {title}
  {\bibinfo {title} {Self-biased tri-state power-multiplexed digital
  metasurface operating at microwave frequencies},\ }\href@noop {} {\bibfield
  {journal} {\bibinfo  {journal} {Opt. Express}\ }\textbf {\bibinfo {volume}
  {28}},\ \bibinfo {pages} {5410} (\bibinfo {year}
  {2020}{\natexlab{b}})}\BibitemShut {NoStop}%
\bibitem [{\citenamefont {Quevedo-Teruel}\ \emph {et~al.}(2019)\citenamefont
  {Quevedo-Teruel}, \citenamefont {Chen}, \citenamefont {D{\'\i}az-Rubio},
  \citenamefont {Gok}, \citenamefont {Grbic}, \citenamefont {Minatti},
  \citenamefont {Martini}, \citenamefont {Maci}, \citenamefont {Eleftheriades},
  \citenamefont {Chen} \emph {et~al.}}]{quevedo2019roadmap}%
  \BibitemOpen
  \bibfield  {author} {\bibinfo {author} {\bibfnamefont {O.}~\bibnamefont
  {Quevedo-Teruel}}, \bibinfo {author} {\bibfnamefont {H.}~\bibnamefont
  {Chen}}, \bibinfo {author} {\bibfnamefont {A.}~\bibnamefont
  {D{\'\i}az-Rubio}}, \bibinfo {author} {\bibfnamefont {G.}~\bibnamefont
  {Gok}}, \bibinfo {author} {\bibfnamefont {A.}~\bibnamefont {Grbic}}, \bibinfo
  {author} {\bibfnamefont {G.}~\bibnamefont {Minatti}}, \bibinfo {author}
  {\bibfnamefont {E.}~\bibnamefont {Martini}}, \bibinfo {author} {\bibfnamefont
  {S.}~\bibnamefont {Maci}}, \bibinfo {author} {\bibfnamefont {G.~V.}\
  \bibnamefont {Eleftheriades}}, \bibinfo {author} {\bibfnamefont
  {M.}~\bibnamefont {Chen}}, \emph {et~al.},\ }\bibfield  {title} {\bibinfo
  {title} {Roadmap on metasurfaces},\ }\href@noop {} {\bibfield  {journal}
  {\bibinfo  {journal} {J. Opt.}\ }\textbf {\bibinfo {volume} {21}},\ \bibinfo
  {pages} {073002} (\bibinfo {year} {2019})}\BibitemShut {NoStop}%
\bibitem [{\citenamefont {Chen}\ and\ \citenamefont
  {Al{\`u}}(2010)}]{chen2010optical}%
  \BibitemOpen
  \bibfield  {author} {\bibinfo {author} {\bibfnamefont {P.-Y.}\ \bibnamefont
  {Chen}}\ and\ \bibinfo {author} {\bibfnamefont {A.}~\bibnamefont {Al{\`u}}},\
  }\bibfield  {title} {\bibinfo {title} {Optical nanoantenna arrays loaded with
  nonlinear materials},\ }\href@noop {} {\bibfield  {journal} {\bibinfo
  {journal} {Phys. Rev. B}\ }\textbf {\bibinfo {volume} {82}},\ \bibinfo
  {pages} {235405} (\bibinfo {year} {2010})}\BibitemShut {NoStop}%
\bibitem [{\citenamefont {Shadrivov}\ \emph {et~al.}(2012)\citenamefont
  {Shadrivov}, \citenamefont {Kapitanova}, \citenamefont {Maslovski},\ and\
  \citenamefont {Kivshar}}]{shadrivov2012metamaterials}%
  \BibitemOpen
  \bibfield  {author} {\bibinfo {author} {\bibfnamefont {I.~V.}\ \bibnamefont
  {Shadrivov}}, \bibinfo {author} {\bibfnamefont {P.~V.}\ \bibnamefont
  {Kapitanova}}, \bibinfo {author} {\bibfnamefont {S.~I.}\ \bibnamefont
  {Maslovski}},\ and\ \bibinfo {author} {\bibfnamefont {Y.~S.}\ \bibnamefont
  {Kivshar}},\ }\bibfield  {title} {\bibinfo {title} {Metamaterials controlled
  with light},\ }\href@noop {} {\bibfield  {journal} {\bibinfo  {journal}
  {Phys. Rev. Lett.}\ }\textbf {\bibinfo {volume} {109}},\ \bibinfo {pages}
  {083902} (\bibinfo {year} {2012})}\BibitemShut {NoStop}%
\bibitem [{\citenamefont {Luo}\ \emph {et~al.}(2015)\citenamefont {Luo},
  \citenamefont {Chen}, \citenamefont {Long}, \citenamefont {Quarfoth},\ and\
  \citenamefont {Sievenpiper}}]{luo2015self}%
  \BibitemOpen
  \bibfield  {author} {\bibinfo {author} {\bibfnamefont {Z.}~\bibnamefont
  {Luo}}, \bibinfo {author} {\bibfnamefont {X.}~\bibnamefont {Chen}}, \bibinfo
  {author} {\bibfnamefont {J.}~\bibnamefont {Long}}, \bibinfo {author}
  {\bibfnamefont {R.}~\bibnamefont {Quarfoth}},\ and\ \bibinfo {author}
  {\bibfnamefont {D.}~\bibnamefont {Sievenpiper}},\ }\bibfield  {title}
  {\bibinfo {title} {Self-focusing of electromagnetic surface waves on a
  nonlinear impedance surface},\ }\href@noop {} {\bibfield  {journal} {\bibinfo
   {journal} {App. Phys. Lett.}\ }\textbf {\bibinfo {volume} {106}},\ \bibinfo
  {pages} {211102} (\bibinfo {year} {2015})}\BibitemShut {NoStop}%
\bibitem [{\citenamefont {Luo}\ \emph {et~al.}(2019)\citenamefont {Luo},
  \citenamefont {Wang}, \citenamefont {Zhang} \emph
  {et~al.}}]{luo2019intensity}%
  \BibitemOpen
  \bibfield  {author} {\bibinfo {author} {\bibfnamefont {Z.}~\bibnamefont
  {Luo}}, \bibinfo {author} {\bibfnamefont {Q.}~\bibnamefont {Wang}}, \bibinfo
  {author} {\bibfnamefont {X.~G.}\ \bibnamefont {Zhang}}, \emph {et~al.},\
  }\bibfield  {title} {\bibinfo {title} {Intensity-dependent metasurface with
  digitally reconfigurable distribution of nonlinearity},\ }\href@noop {}
  {\bibfield  {journal} {\bibinfo  {journal} {Adv. Opt. Mater.}\ }\textbf
  {\bibinfo {volume} {7}},\ \bibinfo {pages} {1900792} (\bibinfo {year}
  {2019})}\BibitemShut {NoStop}%
\bibitem [{\citenamefont {Kim}\ \emph {et~al.}(2020)\citenamefont {Kim},
  \citenamefont {Li}, \citenamefont {Lee},\ and\ \citenamefont
  {Sievenpiper}}]{kim2020active}%
  \BibitemOpen
  \bibfield  {author} {\bibinfo {author} {\bibfnamefont {S.}~\bibnamefont
  {Kim}}, \bibinfo {author} {\bibfnamefont {A.}~\bibnamefont {Li}}, \bibinfo
  {author} {\bibfnamefont {J.}~\bibnamefont {Lee}},\ and\ \bibinfo {author}
  {\bibfnamefont {D.~F.}\ \bibnamefont {Sievenpiper}},\ }\bibfield  {title}
  {\bibinfo {title} {Active self-tuning metasurface with enhanced absorbing
  frequency range for suppression of high-power surface currents},\ }\href@noop
  {} {\bibfield  {journal} {\bibinfo  {journal} {IEEE Trans. Antennas Propag.}\
  }\textbf {\bibinfo {volume} {69}},\ \bibinfo {pages} {2759} (\bibinfo {year}
  {2020})}\BibitemShut {NoStop}%
\bibitem [{\citenamefont {Li}\ \emph {et~al.}(2017{\natexlab{b}})\citenamefont
  {Li}, \citenamefont {Kim}, \citenamefont {Luo}, \citenamefont {Li},
  \citenamefont {Long},\ and\ \citenamefont {Sievenpiper}}]{li2017high}%
  \BibitemOpen
  \bibfield  {author} {\bibinfo {author} {\bibfnamefont {A.}~\bibnamefont
  {Li}}, \bibinfo {author} {\bibfnamefont {S.}~\bibnamefont {Kim}}, \bibinfo
  {author} {\bibfnamefont {Y.}~\bibnamefont {Luo}}, \bibinfo {author}
  {\bibfnamefont {Y.}~\bibnamefont {Li}}, \bibinfo {author} {\bibfnamefont
  {J.}~\bibnamefont {Long}},\ and\ \bibinfo {author} {\bibfnamefont {D.~F.}\
  \bibnamefont {Sievenpiper}},\ }\bibfield  {title} {\bibinfo {title}
  {High-power transistor-based tunable and switchable metasurface absorber},\
  }\href@noop {} {\bibfield  {journal} {\bibinfo  {journal} {IEEE Trans.
  Microw. Theory Tech.}\ }\textbf {\bibinfo {volume} {65}},\ \bibinfo {pages}
  {2810} (\bibinfo {year} {2017}{\natexlab{b}})}\BibitemShut {NoStop}%
\bibitem [{\citenamefont {Luo}\ \emph {et~al.}(2022)\citenamefont {Luo},
  \citenamefont {Ren}, \citenamefont {Zhou}, \citenamefont {Chen},
  \citenamefont {Cheng}, \citenamefont {Ma},\ and\ \citenamefont
  {Cui}}]{luo2022high}%
  \BibitemOpen
  \bibfield  {author} {\bibinfo {author} {\bibfnamefont {Z.}~\bibnamefont
  {Luo}}, \bibinfo {author} {\bibfnamefont {X.}~\bibnamefont {Ren}}, \bibinfo
  {author} {\bibfnamefont {L.}~\bibnamefont {Zhou}}, \bibinfo {author}
  {\bibfnamefont {Y.}~\bibnamefont {Chen}}, \bibinfo {author} {\bibfnamefont
  {Q.}~\bibnamefont {Cheng}}, \bibinfo {author} {\bibfnamefont {H.~F.}\
  \bibnamefont {Ma}},\ and\ \bibinfo {author} {\bibfnamefont {T.~J.}\
  \bibnamefont {Cui}},\ }\bibfield  {title} {\bibinfo {title} {A
  high-performance nonlinear metasurface for spatial-wave absorption},\
  }\href@noop {} {\bibfield  {journal} {\bibinfo  {journal} {Adv. Funct.
  Mater.}\ }\textbf {\bibinfo {volume} {32}},\ \bibinfo {pages} {2109544}
  (\bibinfo {year} {2022})}\BibitemShut {NoStop}%
\bibitem [{\citenamefont {Homma}\ \emph {et~al.}(2022)\citenamefont {Homma},
  \citenamefont {Akram}, \citenamefont {Fathnan}, \citenamefont {Lee},
  \citenamefont {Christopoulos},\ and\ \citenamefont
  {Wakatsuchi}}]{homma2022anisotropic}%
  \BibitemOpen
  \bibfield  {author} {\bibinfo {author} {\bibfnamefont {H.}~\bibnamefont
  {Homma}}, \bibinfo {author} {\bibfnamefont {M.~R.}\ \bibnamefont {Akram}},
  \bibinfo {author} {\bibfnamefont {A.~A.}\ \bibnamefont {Fathnan}}, \bibinfo
  {author} {\bibfnamefont {J.}~\bibnamefont {Lee}}, \bibinfo {author}
  {\bibfnamefont {C.}~\bibnamefont {Christopoulos}},\ and\ \bibinfo {author}
  {\bibfnamefont {H.}~\bibnamefont {Wakatsuchi}},\ }\bibfield  {title}
  {\bibinfo {title} {Anisotropic impedance surfaces activated by incident
  waveform},\ }\href@noop {} {\bibfield  {journal} {\bibinfo  {journal}
  {Nanophotonics}\ }\textbf {\bibinfo {volume} {11}},\ \bibinfo {pages} {1989}
  (\bibinfo {year} {2022})}\BibitemShut {NoStop}%
\bibitem [{\citenamefont {Wakatsuchi}\ \emph {et~al.}(2015)\citenamefont
  {Wakatsuchi}, \citenamefont {Anzai}, \citenamefont {Rushton}, \citenamefont
  {Gao}, \citenamefont {Kim},\ and\ \citenamefont
  {Sievenpiper}}]{wakatsuchi2015waveform}%
  \BibitemOpen
  \bibfield  {author} {\bibinfo {author} {\bibfnamefont {H.}~\bibnamefont
  {Wakatsuchi}}, \bibinfo {author} {\bibfnamefont {D.}~\bibnamefont {Anzai}},
  \bibinfo {author} {\bibfnamefont {J.~J.}\ \bibnamefont {Rushton}}, \bibinfo
  {author} {\bibfnamefont {F.}~\bibnamefont {Gao}}, \bibinfo {author}
  {\bibfnamefont {S.}~\bibnamefont {Kim}},\ and\ \bibinfo {author}
  {\bibfnamefont {D.~F.}\ \bibnamefont {Sievenpiper}},\ }\bibfield  {title}
  {\bibinfo {title} {Waveform selectivity at the same frequency},\ }\href@noop
  {} {\bibfield  {journal} {\bibinfo  {journal} {Sci. Rep.}\ }\textbf {\bibinfo
  {volume} {5}},\ \bibinfo {pages} {1} (\bibinfo {year} {2015})}\BibitemShut
  {NoStop}%
\bibitem [{\citenamefont {Wakatsuchi}\ \emph
  {et~al.}(2013{\natexlab{b}})\citenamefont {Wakatsuchi}, \citenamefont {Kim},
  \citenamefont {Rushton},\ and\ \citenamefont
  {Sievenpiper}}]{15Wakatsuchi:2013}%
  \BibitemOpen
  \bibfield  {author} {\bibinfo {author} {\bibfnamefont {H.}~\bibnamefont
  {Wakatsuchi}}, \bibinfo {author} {\bibfnamefont {S.}~\bibnamefont {Kim}},
  \bibinfo {author} {\bibfnamefont {J.~J.}\ \bibnamefont {Rushton}},\ and\
  \bibinfo {author} {\bibfnamefont {D.~F.}\ \bibnamefont {Sievenpiper}},\
  }\bibfield  {title} {\bibinfo {title} {Waveform-dependent absorbing
  metasurfaces},\ }\href@noop {} {\bibfield  {journal} {\bibinfo  {journal}
  {Phy. Rev. Lett.}\ }\textbf {\bibinfo {volume} {111}},\ \bibinfo {pages}
  {245501} (\bibinfo {year} {2013}{\natexlab{b}})}\BibitemShut {NoStop}%
\bibitem [{\citenamefont {Wakatsuchi}\ \emph {et~al.}(2019)\citenamefont
  {Wakatsuchi}, \citenamefont {Long},\ and\ \citenamefont
  {Sievenpiper}}]{14Wakatsuchi:2019}%
  \BibitemOpen
  \bibfield  {author} {\bibinfo {author} {\bibfnamefont {H.}~\bibnamefont
  {Wakatsuchi}}, \bibinfo {author} {\bibfnamefont {J.}~\bibnamefont {Long}},\
  and\ \bibinfo {author} {\bibfnamefont {D.~F.}\ \bibnamefont {Sievenpiper}},\
  }\bibfield  {title} {\bibinfo {title} {Waveform selective surfaces},\
  }\href@noop {} {\bibfield  {journal} {\bibinfo  {journal} {Adv. Funct.
  Mater.}\ }\textbf {\bibinfo {volume} {29}},\ \bibinfo {pages} {1806386}
  (\bibinfo {year} {2019})}\BibitemShut {NoStop}%
\bibitem [{\citenamefont {Ushikoshi}\ \emph {et~al.}(2020)\citenamefont
  {Ushikoshi}, \citenamefont {Tanikawa}, \citenamefont {Asano}, \citenamefont
  {Sanji}, \citenamefont {Ikeda}, \citenamefont {Anzai},\ and\ \citenamefont
  {Wakatsuchi}}]{ushikoshi2020experimental}%
  \BibitemOpen
  \bibfield  {author} {\bibinfo {author} {\bibfnamefont {D.}~\bibnamefont
  {Ushikoshi}}, \bibinfo {author} {\bibfnamefont {M.}~\bibnamefont {Tanikawa}},
  \bibinfo {author} {\bibfnamefont {K.}~\bibnamefont {Asano}}, \bibinfo
  {author} {\bibfnamefont {K.}~\bibnamefont {Sanji}}, \bibinfo {author}
  {\bibfnamefont {M.}~\bibnamefont {Ikeda}}, \bibinfo {author} {\bibfnamefont
  {D.}~\bibnamefont {Anzai}},\ and\ \bibinfo {author} {\bibfnamefont
  {H.}~\bibnamefont {Wakatsuchi}},\ }\bibfield  {title} {\bibinfo {title}
  {Experimental demonstration of waveform-selective metasurface varying
  wireless communication characteristics at the same frequency band of 2.4
  ghz},\ }\href@noop {} {\bibfield  {journal} {\bibinfo  {journal} {Electron.
  Lett.}\ }\textbf {\bibinfo {volume} {56}},\ \bibinfo {pages} {160} (\bibinfo
  {year} {2020})}\BibitemShut {NoStop}%
\bibitem [{\citenamefont {Vellucci}\ \emph
  {et~al.}(2019{\natexlab{a}})\citenamefont {Vellucci}, \citenamefont {Monti},
  \citenamefont {Barbuto}, \citenamefont {Toscano},\ and\ \citenamefont
  {Bilotti}}]{kVellucci:2020}%
  \BibitemOpen
  \bibfield  {author} {\bibinfo {author} {\bibfnamefont {S.}~\bibnamefont
  {Vellucci}}, \bibinfo {author} {\bibfnamefont {A.}~\bibnamefont {Monti}},
  \bibinfo {author} {\bibfnamefont {M.}~\bibnamefont {Barbuto}}, \bibinfo
  {author} {\bibfnamefont {A.}~\bibnamefont {Toscano}},\ and\ \bibinfo {author}
  {\bibfnamefont {F.}~\bibnamefont {Bilotti}},\ }\bibfield  {title} {\bibinfo
  {title} {Waveform-selective mantle cloaks for intelligent antennas},\
  }\href@noop {} {\bibfield  {journal} {\bibinfo  {journal} {IEEE Trans.
  Antennas Propag.}\ }\textbf {\bibinfo {volume} {68}},\ \bibinfo {pages}
  {1717} (\bibinfo {year} {2019}{\natexlab{a}})}\BibitemShut {NoStop}%
\bibitem [{\citenamefont {Wakatsuchi}(2015)}]{wakatsuchi2015time}%
  \BibitemOpen
  \bibfield  {author} {\bibinfo {author} {\bibfnamefont {H.}~\bibnamefont
  {Wakatsuchi}},\ }\bibfield  {title} {\bibinfo {title} {Time-domain filtering
  of metasurfaces},\ }\href@noop {} {\bibfield  {journal} {\bibinfo  {journal}
  {Sci. Rep.}\ }\textbf {\bibinfo {volume} {5}},\ \bibinfo {pages} {1}
  (\bibinfo {year} {2015})}\BibitemShut {NoStop}%
\bibitem [{\citenamefont {Vellucci}\ \emph
  {et~al.}(2019{\natexlab{b}})\citenamefont {Vellucci}, \citenamefont {Monti},
  \citenamefont {Barbuto}, \citenamefont {Toscano},\ and\ \citenamefont
  {Bilotti}}]{vellucci2019waveform}%
  \BibitemOpen
  \bibfield  {author} {\bibinfo {author} {\bibfnamefont {S.}~\bibnamefont
  {Vellucci}}, \bibinfo {author} {\bibfnamefont {A.}~\bibnamefont {Monti}},
  \bibinfo {author} {\bibfnamefont {M.}~\bibnamefont {Barbuto}}, \bibinfo
  {author} {\bibfnamefont {A.}~\bibnamefont {Toscano}},\ and\ \bibinfo {author}
  {\bibfnamefont {F.}~\bibnamefont {Bilotti}},\ }\bibfield  {title} {\bibinfo
  {title} {Waveform-selective mantle cloaks for intelligent antennas},\
  }\href@noop {} {\bibfield  {journal} {\bibinfo  {journal} {IEEE Trans.
  Antennas Propag.}\ }\textbf {\bibinfo {volume} {68}},\ \bibinfo {pages}
  {1717} (\bibinfo {year} {2019}{\natexlab{b}})}\BibitemShut {NoStop}%
\bibitem [{\citenamefont {Asano}\ \emph {et~al.}(2020)\citenamefont {Asano},
  \citenamefont {Nakasha},\ and\ \citenamefont
  {Wakatsuchi}}]{asano2020simplified}%
  \BibitemOpen
  \bibfield  {author} {\bibinfo {author} {\bibfnamefont {K.}~\bibnamefont
  {Asano}}, \bibinfo {author} {\bibfnamefont {T.}~\bibnamefont {Nakasha}},\
  and\ \bibinfo {author} {\bibfnamefont {H.}~\bibnamefont {Wakatsuchi}},\
  }\bibfield  {title} {\bibinfo {title} {Simplified equivalent circuit approach
  for designing time-domain responses of waveform-selective metasurfaces},\
  }\href@noop {} {\bibfield  {journal} {\bibinfo  {journal} {App. Phys. Lett.}\
  }\textbf {\bibinfo {volume} {116}},\ \bibinfo {pages} {171603} (\bibinfo
  {year} {2020})}\BibitemShut {NoStop}%
\bibitem [{\citenamefont {Holloway}\ \emph {et~al.}(2005)\citenamefont
  {Holloway}, \citenamefont {Mohamed}, \citenamefont {Kuester},\ and\
  \citenamefont {Dienstfrey}}]{holloway2005reflection}%
  \BibitemOpen
  \bibfield  {author} {\bibinfo {author} {\bibfnamefont {C.~L.}\ \bibnamefont
  {Holloway}}, \bibinfo {author} {\bibfnamefont {M.~A.}\ \bibnamefont
  {Mohamed}}, \bibinfo {author} {\bibfnamefont {E.~F.}\ \bibnamefont
  {Kuester}},\ and\ \bibinfo {author} {\bibfnamefont {A.}~\bibnamefont
  {Dienstfrey}},\ }\bibfield  {title} {\bibinfo {title} {Reflection and
  transmission properties of a metafilm: With an application to a controllable
  surface composed of resonant particles},\ }\href@noop {} {\bibfield
  {journal} {\bibinfo  {journal} {IEEE Trans. Electromagn. Compat.}\ }\textbf
  {\bibinfo {volume} {47}},\ \bibinfo {pages} {853} (\bibinfo {year}
  {2005})}\BibitemShut {NoStop}%
\bibitem [{\citenamefont {Fowles}(1989)}]{fowles1989introduction}%
  \BibitemOpen
  \bibfield  {author} {\bibinfo {author} {\bibfnamefont {G.~R.}\ \bibnamefont
  {Fowles}},\ }\href@noop {} {\emph {\bibinfo {title} {Introduction to modern
  optics}}}\ (\bibinfo  {publisher} {Dover Publications, New York},\ \bibinfo
  {year} {1989})\BibitemShut {NoStop}%
\bibitem [{\citenamefont {Shams}\ \emph {et~al.}(2017)\citenamefont {Shams},
  \citenamefont {Jiang}, \citenamefont {Rahman}, \citenamefont {Cheng},
  \citenamefont {Hesler}, \citenamefont {Fay},\ and\ \citenamefont
  {Liu}}]{shams2017740}%
  \BibitemOpen
  \bibfield  {author} {\bibinfo {author} {\bibfnamefont {M.~I.~B.}\
  \bibnamefont {Shams}}, \bibinfo {author} {\bibfnamefont {Z.}~\bibnamefont
  {Jiang}}, \bibinfo {author} {\bibfnamefont {S.~M.}\ \bibnamefont {Rahman}},
  \bibinfo {author} {\bibfnamefont {L.-J.}\ \bibnamefont {Cheng}}, \bibinfo
  {author} {\bibfnamefont {J.~L.}\ \bibnamefont {Hesler}}, \bibinfo {author}
  {\bibfnamefont {P.}~\bibnamefont {Fay}},\ and\ \bibinfo {author}
  {\bibfnamefont {L.}~\bibnamefont {Liu}},\ }\bibfield  {title} {\bibinfo
  {title} {A 740-ghz dynamic two-dimensional beam-steering and forming antenna
  based on photo-induced reconfigurable fresnel zone plates},\ }\href@noop {}
  {\bibfield  {journal} {\bibinfo  {journal} {IEEE Trans. Terahertz Sci.
  Technol.}\ }\textbf {\bibinfo {volume} {7}},\ \bibinfo {pages} {310}
  (\bibinfo {year} {2017})}\BibitemShut {NoStop}%
\bibitem [{\citenamefont {Fathnan}\ \emph
  {et~al.}(2020{\natexlab{a}})\citenamefont {Fathnan}, \citenamefont {Liu},\
  and\ \citenamefont {Powell}}]{fathnan2020achromatic}%
  \BibitemOpen
  \bibfield  {author} {\bibinfo {author} {\bibfnamefont {A.~A.}\ \bibnamefont
  {Fathnan}}, \bibinfo {author} {\bibfnamefont {M.}~\bibnamefont {Liu}},\ and\
  \bibinfo {author} {\bibfnamefont {D.~A.}\ \bibnamefont {Powell}},\ }\bibfield
   {title} {\bibinfo {title} {Achromatic huygens’ metalenses with deeply
  subwavelength thickness},\ }\href@noop {} {\bibfield  {journal} {\bibinfo
  {journal} {Adv. Opt. Mater.}\ }\textbf {\bibinfo {volume} {8}},\ \bibinfo
  {pages} {2000754} (\bibinfo {year} {2020}{\natexlab{a}})}\BibitemShut
  {NoStop}%
\bibitem [{\citenamefont {Epstein}\ and\ \citenamefont
  {Eleftheriades}(2016)}]{epstein2016huygens}%
  \BibitemOpen
  \bibfield  {author} {\bibinfo {author} {\bibfnamefont {A.}~\bibnamefont
  {Epstein}}\ and\ \bibinfo {author} {\bibfnamefont {G.~V.}\ \bibnamefont
  {Eleftheriades}},\ }\bibfield  {title} {\bibinfo {title} {Huygens’
  metasurfaces via the equivalence principle: design and applications},\
  }\href@noop {} {\bibfield  {journal} {\bibinfo  {journal} {JOSA B}\ }\textbf
  {\bibinfo {volume} {33}},\ \bibinfo {pages} {A31} (\bibinfo {year}
  {2016})}\BibitemShut {NoStop}%
\bibitem [{\citenamefont {Monticone}\ \emph {et~al.}(2013)\citenamefont
  {Monticone}, \citenamefont {Estakhri},\ and\ \citenamefont
  {Alu}}]{monticone2013full}%
  \BibitemOpen
  \bibfield  {author} {\bibinfo {author} {\bibfnamefont {F.}~\bibnamefont
  {Monticone}}, \bibinfo {author} {\bibfnamefont {N.~M.}\ \bibnamefont
  {Estakhri}},\ and\ \bibinfo {author} {\bibfnamefont {A.}~\bibnamefont
  {Alu}},\ }\bibfield  {title} {\bibinfo {title} {Full control of nanoscale
  optical transmission with a composite metascreen},\ }\href@noop {} {\bibfield
   {journal} {\bibinfo  {journal} {Phys. Rev. Lett.}\ }\textbf {\bibinfo
  {volume} {110}},\ \bibinfo {pages} {203903} (\bibinfo {year}
  {2013})}\BibitemShut {NoStop}%
\bibitem [{\citenamefont {Decker}\ \emph {et~al.}(2015)\citenamefont {Decker},
  \citenamefont {Staude}, \citenamefont {Falkner}, \citenamefont {Dominguez},
  \citenamefont {Neshev}, \citenamefont {Brener}, \citenamefont {Pertsch},\
  and\ \citenamefont {Kivshar}}]{decker2015high}%
  \BibitemOpen
  \bibfield  {author} {\bibinfo {author} {\bibfnamefont {M.}~\bibnamefont
  {Decker}}, \bibinfo {author} {\bibfnamefont {I.}~\bibnamefont {Staude}},
  \bibinfo {author} {\bibfnamefont {M.}~\bibnamefont {Falkner}}, \bibinfo
  {author} {\bibfnamefont {J.}~\bibnamefont {Dominguez}}, \bibinfo {author}
  {\bibfnamefont {D.~N.}\ \bibnamefont {Neshev}}, \bibinfo {author}
  {\bibfnamefont {I.}~\bibnamefont {Brener}}, \bibinfo {author} {\bibfnamefont
  {T.}~\bibnamefont {Pertsch}},\ and\ \bibinfo {author} {\bibfnamefont {Y.~S.}\
  \bibnamefont {Kivshar}},\ }\bibfield  {title} {\bibinfo {title}
  {High-efficiency dielectric huygens’ surfaces},\ }\href@noop {} {\bibfield
  {journal} {\bibinfo  {journal} {Adv. Opt. Mater.}\ }\textbf {\bibinfo
  {volume} {3}},\ \bibinfo {pages} {813} (\bibinfo {year} {2015})}\BibitemShut
  {NoStop}%
\bibitem [{\citenamefont {Boashash}(2015)}]{boashash2015time}%
  \BibitemOpen
  \bibfield  {author} {\bibinfo {author} {\bibfnamefont {B.}~\bibnamefont
  {Boashash}},\ }\href@noop {} {\emph {\bibinfo {title} {Time-frequency signal
  analysis and processing: a comprehensive reference}}}\ (\bibinfo  {publisher}
  {Academic press, London},\ \bibinfo {year} {2015})\BibitemShut {NoStop}%
\bibitem [{\citenamefont {Sejdi{\'c}}\ \emph {et~al.}(2009)\citenamefont
  {Sejdi{\'c}}, \citenamefont {Djurovi{\'c}},\ and\ \citenamefont
  {Jiang}}]{sejdic2009time}%
  \BibitemOpen
  \bibfield  {author} {\bibinfo {author} {\bibfnamefont {E.}~\bibnamefont
  {Sejdi{\'c}}}, \bibinfo {author} {\bibfnamefont {I.}~\bibnamefont
  {Djurovi{\'c}}},\ and\ \bibinfo {author} {\bibfnamefont {J.}~\bibnamefont
  {Jiang}},\ }\bibfield  {title} {\bibinfo {title} {Time--frequency feature
  representation using energy concentration: An overview of recent advances},\
  }\href@noop {} {\bibfield  {journal} {\bibinfo  {journal} {Digit. Signal
  Process.}\ }\textbf {\bibinfo {volume} {19}},\ \bibinfo {pages} {153}
  (\bibinfo {year} {2009})}\BibitemShut {NoStop}%
\bibitem [{\citenamefont {Fathnan}\ \emph
  {et~al.}(2020{\natexlab{b}})\citenamefont {Fathnan}, \citenamefont {Olk},\
  and\ \citenamefont {Powell}}]{fathnan2020bandwidth}%
  \BibitemOpen
  \bibfield  {author} {\bibinfo {author} {\bibfnamefont {A.~A.}\ \bibnamefont
  {Fathnan}}, \bibinfo {author} {\bibfnamefont {A.~E.}\ \bibnamefont {Olk}},\
  and\ \bibinfo {author} {\bibfnamefont {D.~A.}\ \bibnamefont {Powell}},\
  }\bibfield  {title} {\bibinfo {title} {Bandwidth limit and synthesis approach
  for single resonance ultrathin metasurfaces},\ }\href@noop {} {\bibfield
  {journal} {\bibinfo  {journal} {J. Phys. D: Appl. Phys.}\ }\textbf {\bibinfo
  {volume} {53}},\ \bibinfo {pages} {495304} (\bibinfo {year}
  {2020}{\natexlab{b}})}\BibitemShut {NoStop}%
\bibitem [{\citenamefont {Olk}\ and\ \citenamefont
  {Powell}(2019)}]{olk2019accurate}%
  \BibitemOpen
  \bibfield  {author} {\bibinfo {author} {\bibfnamefont {A.~E.}\ \bibnamefont
  {Olk}}\ and\ \bibinfo {author} {\bibfnamefont {D.~A.}\ \bibnamefont
  {Powell}},\ }\bibfield  {title} {\bibinfo {title} {Accurate metasurface
  synthesis incorporating near-field coupling effects},\ }\href@noop {}
  {\bibfield  {journal} {\bibinfo  {journal} {Phys. Rev. App.}\ }\textbf
  {\bibinfo {volume} {11}},\ \bibinfo {pages} {064007} (\bibinfo {year}
  {2019})}\BibitemShut {NoStop}%
\bibitem [{\citenamefont {Wu}\ and\ \citenamefont
  {Zhang}(2019)}]{wu2019intelligent}%
  \BibitemOpen
  \bibfield  {author} {\bibinfo {author} {\bibfnamefont {Q.}~\bibnamefont
  {Wu}}\ and\ \bibinfo {author} {\bibfnamefont {R.}~\bibnamefont {Zhang}},\
  }\bibfield  {title} {\bibinfo {title} {Intelligent reflecting surface
  enhanced wireless network via joint active and passive beamforming},\
  }\href@noop {} {\bibfield  {journal} {\bibinfo  {journal} {IEEE Trans. Wirel.
  Commun.}\ }\textbf {\bibinfo {volume} {18}},\ \bibinfo {pages} {5394}
  (\bibinfo {year} {2019})}\BibitemShut {NoStop}%
\bibitem [{\citenamefont {{\"O}zdogan}\ \emph {et~al.}(2019)\citenamefont
  {{\"O}zdogan}, \citenamefont {Bj{\"o}rnson},\ and\ \citenamefont
  {Larsson}}]{ozdogan2019intelligent}%
  \BibitemOpen
  \bibfield  {author} {\bibinfo {author} {\bibfnamefont {{\"O}.}~\bibnamefont
  {{\"O}zdogan}}, \bibinfo {author} {\bibfnamefont {E.}~\bibnamefont
  {Bj{\"o}rnson}},\ and\ \bibinfo {author} {\bibfnamefont {E.~G.}\ \bibnamefont
  {Larsson}},\ }\bibfield  {title} {\bibinfo {title} {Intelligent reflecting
  surfaces: Physics, propagation, and pathloss modeling},\ }\href@noop {}
  {\bibfield  {journal} {\bibinfo  {journal} {IEEE Wirel. Commun.}\ }\textbf
  {\bibinfo {volume} {9}},\ \bibinfo {pages} {581} (\bibinfo {year}
  {2019})}\BibitemShut {NoStop}%
\bibitem [{\citenamefont {Sugiura}\ \emph {et~al.}(2021)\citenamefont
  {Sugiura}, \citenamefont {Kawai}, \citenamefont {Matsui}, \citenamefont
  {Lee},\ and\ \citenamefont {Iizuka}}]{sugiura2021joint}%
  \BibitemOpen
  \bibfield  {author} {\bibinfo {author} {\bibfnamefont {S.}~\bibnamefont
  {Sugiura}}, \bibinfo {author} {\bibfnamefont {Y.}~\bibnamefont {Kawai}},
  \bibinfo {author} {\bibfnamefont {T.}~\bibnamefont {Matsui}}, \bibinfo
  {author} {\bibfnamefont {T.}~\bibnamefont {Lee}},\ and\ \bibinfo {author}
  {\bibfnamefont {H.}~\bibnamefont {Iizuka}},\ }\bibfield  {title} {\bibinfo
  {title} {Joint beam and polarization forming of intelligent reflecting
  surfaces for wireless communications},\ }\href@noop {} {\bibfield  {journal}
  {\bibinfo  {journal} {IEEE Trans. Veh. Technol.}\ }\textbf {\bibinfo {volume}
  {70}},\ \bibinfo {pages} {1648} (\bibinfo {year} {2021})}\BibitemShut
  {NoStop}%
\bibitem [{\citenamefont {Dai}\ \emph {et~al.}(2020)\citenamefont {Dai},
  \citenamefont {Wang}, \citenamefont {Wang}, \citenamefont {Yang},
  \citenamefont {Tan}, \citenamefont {Bi}, \citenamefont {Xu}, \citenamefont
  {Yang}, \citenamefont {Chen}, \citenamefont {Di~Renzo} \emph
  {et~al.}}]{dai2020reconfigurable}%
  \BibitemOpen
  \bibfield  {author} {\bibinfo {author} {\bibfnamefont {L.}~\bibnamefont
  {Dai}}, \bibinfo {author} {\bibfnamefont {B.}~\bibnamefont {Wang}}, \bibinfo
  {author} {\bibfnamefont {M.}~\bibnamefont {Wang}}, \bibinfo {author}
  {\bibfnamefont {X.}~\bibnamefont {Yang}}, \bibinfo {author} {\bibfnamefont
  {J.}~\bibnamefont {Tan}}, \bibinfo {author} {\bibfnamefont {S.}~\bibnamefont
  {Bi}}, \bibinfo {author} {\bibfnamefont {S.}~\bibnamefont {Xu}}, \bibinfo
  {author} {\bibfnamefont {F.}~\bibnamefont {Yang}}, \bibinfo {author}
  {\bibfnamefont {Z.}~\bibnamefont {Chen}}, \bibinfo {author} {\bibfnamefont
  {M.}~\bibnamefont {Di~Renzo}}, \emph {et~al.},\ }\bibfield  {title} {\bibinfo
  {title} {Reconfigurable intelligent surface-based wireless communications:
  Antenna design, prototyping, and experimental results},\ }\href@noop {}
  {\bibfield  {journal} {\bibinfo  {journal} {IEEE Access}\ }\textbf {\bibinfo
  {volume} {8}},\ \bibinfo {pages} {45913} (\bibinfo {year}
  {2020})}\BibitemShut {NoStop}%
\bibitem [{\citenamefont {Zhang}\ \emph {et~al.}(2018)\citenamefont {Zhang},
  \citenamefont {Chen}, \citenamefont {Liu}, \citenamefont {Zhang},
  \citenamefont {Zhao}, \citenamefont {Dai}, \citenamefont {Bai}, \citenamefont
  {Wan}, \citenamefont {Cheng}, \citenamefont {Castaldi} \emph
  {et~al.}}]{zhang2018space}%
  \BibitemOpen
  \bibfield  {author} {\bibinfo {author} {\bibfnamefont {L.}~\bibnamefont
  {Zhang}}, \bibinfo {author} {\bibfnamefont {X.~Q.}\ \bibnamefont {Chen}},
  \bibinfo {author} {\bibfnamefont {S.}~\bibnamefont {Liu}}, \bibinfo {author}
  {\bibfnamefont {Q.}~\bibnamefont {Zhang}}, \bibinfo {author} {\bibfnamefont
  {J.}~\bibnamefont {Zhao}}, \bibinfo {author} {\bibfnamefont {J.~Y.}\
  \bibnamefont {Dai}}, \bibinfo {author} {\bibfnamefont {G.~D.}\ \bibnamefont
  {Bai}}, \bibinfo {author} {\bibfnamefont {X.}~\bibnamefont {Wan}}, \bibinfo
  {author} {\bibfnamefont {Q.}~\bibnamefont {Cheng}}, \bibinfo {author}
  {\bibfnamefont {G.}~\bibnamefont {Castaldi}}, \emph {et~al.},\ }\bibfield
  {title} {\bibinfo {title} {Space-time-coding digital metasurfaces},\
  }\href@noop {} {\bibfield  {journal} {\bibinfo  {journal} {Nat. Commun.}\
  }\textbf {\bibinfo {volume} {9}},\ \bibinfo {pages} {1} (\bibinfo {year}
  {2018})}\BibitemShut {NoStop}%
\bibitem [{\citenamefont {Pozar}(1988)}]{pozar2011microwave}%
  \BibitemOpen
  \bibfield  {author} {\bibinfo {author} {\bibfnamefont {D.~M.}\ \bibnamefont
  {Pozar}},\ }\href@noop {} {\emph {\bibinfo {title} {Microwave engineering}}}\
  (\bibinfo  {publisher} {Wiley, New York},\ \bibinfo {year}
  {1988})\BibitemShut {NoStop}%
\end{thebibliography}%
\end{document}